\documentclass[aps,twocolumn,showpacs,pre]{revtex4-1}
\usepackage{graphicx,graphics}
\usepackage{amsmath,amssymb,amsfonts}
\usepackage{multirow}
\usepackage{hyperref}
\usepackage{graphicx}

\begin{document}
\title{Aging in transport processes on networks with stochastic cumulative damage}
\author{A.P. Riascos}
\email{aperezr@fisica.unam.mx}
\affiliation{Instituto de F\'isica, Universidad Nacional Aut\'onoma de M\'exico, 
Apartado Postal 20-364, 01000 Ciudad de M\'exico, M\'exico}

\author{J. Wang-Michelitsch}
\affiliation{Independent researcher, Paris, France}
\author{T.M. Michelitsch} 
\affiliation{Sorbonne Universit\'e, Institut Jean le Rond d'Alembert, CNRS UMR 7190,4 place Jussieu, 75252 Paris cedex 05, France}

\date{\today}

\begin{abstract}
In this paper we explore the evolution of transport capacity on networks with stochastic incidence of damage and accumulation of faults in their connections. For each damaged configuration of the network, we analyze a Markovian random walker that hops over weighted links that quantify the capacity of transport of each connection. The weights of the links in the network evolve due to randomly occurring damage effects that reduce gradually the transport capacity of the structure. We introduce a global measure to determine the functionality of each configuration and how the system ages due to the accumulation of damage that cannot be repaired completely. Then, by assuming a minimum value of the functionality required for the system to be ``alive'', we explore the statistics of the lifetimes for several realizations of this process in different types of networks. Finally, we analyze the characteristic longevity of such a system and its relation with the ``complexity''  of the network structure. One finding is that systems with greater complexity live longer.  Our approach introduces a model of aging processes relating the reduction of functionality with the accumulation of ``misrepairs'' and the lifetime of a complex system.
\end{abstract}


\maketitle

\section{Introduction}
\label{intro}
One of the main features observed in many ``complex systems'' such as 
living organisms \cite{Kirkwood2005,LopezOtin2013,Cohen2016,Farrell2016,Taneja2016}, 
social systems \cite{Manrique2018}, 
corporations, and civilizations \cite{Hershey1986}, is that these systems 
exhibit aging and a limited lifespan \cite{west2018scale}. 
These systems are continuously subjected to external damage exposure. 
In order to ``survive,'' the system needs continuously respond 
to damage impact with ``reparation processes'' maintaining both immediate and long-term survival. 
The reparation process in a complex system is performed under a time constraint that 
the functionality of the entire system during the reparation has to be maintained. 
Due to this time constraint of the reparation process, when the damage impact is ``too severe'', 
the system is not able to re-establish the 
original undamaged structure but generates a repaired structure 
with altered properties a so-called ``{\it misrepar}.'' Simply speaking the misrepair mechanism can be 
considered as a compromise between two needs: the reparation as good as possible and the reparation as fast as necessary. 
In many cases, the alteration in a misrepaired structure compared to the initial 
undamaged structure may be very `small' and the misrepair may be even `close' to perfect reparation \cite{WangMiWun2009}.
\\[2mm]
Therefore, the mechanism of misrepair guarantees the immediate survival as a result of a 
reparation process that takes place sufficiently fast in order to continuously maintain 
functionality and avoiding fatal damage
consequences. 
However, the price to be paid of this fast reparation process is a misrepaired altered structure 
with reduced functionality 
compared to the initial undamaged structure. 
\\[2mm]
Now imagine a living being is continuously exposed to damage impact where as a result continuously misrepairs are generated. As a consequence gradual accumulation of misrepairs is taking place in the organism 
deteriorating gradually its functionality \cite{Kirkwood2005,WangMiWun2009}. 
The {\it accumulation of misrepairs} has been suggested to explain several 
{\it aging} phenomena in living beings and it has been 
conjectured that these mechanisms might hold in a wider class of certain complex systems 
\cite{WangMiWun2009,WangMi2018,WangMi2015}.
The occurrence of misrepair and as a consequence aging hence are {\it necessary} 
for immediate survival. Namely,
without aging, there would be no immediate survival upon damage impact, 
and blocking the aging process would mean blocking the reparation 
responses in the complex system. Therefore, aging and final death are the inevitable prizes 
to be paid to have a certain finite lifespan \cite{WangMiWun2009,WangMi2015}. 
\\[2mm]
In this paper, we model the aging process of a complex system that is governed by an ``accumulation of misrepairs'' mechanism. We emphasize that the term ``accumulation'' is not to be understood in a linear sense i.e. as a simple superposition of `misrepaired' structures. 
We describe the complex system in the present paper as an undirected connected weighted network. 
Due to the lack of a precise definition of the notion of {\it complexity}, 
we compare the aging process in different kinds of networks with different topologies allowing 
to distinguish their complexities qualitatively. We assume the complex system to be ``alive'' when all parts of the network can communicate with each other in a sufficiently short time. We describe the ability of communication of the network by its transport capacity modeled by a random walker that navigates through the network.  We assume that the complex system is alive when a global quantity that describes the transport capability of the structure is greater 
than a certain critical threshold. Then the communication in the system is assumed to be fast enough to guarantee functionality to
operate. In the initial configuration of the system, we assume the network to be a ``perfect structure'' described by an undirected connected network, and this structure is gradually altered by progressing accumulation of misrepairs.
One main outcome of our model is that increased complexity increases the lifespan of a system. 
\section{Transport on networks with cumulative damage}
\label{SectionModel}
In this section, we introduce a phenomenological model for aging processes in a complex system represented 
by a network. The model relies on three characteristics: (1) We consider a network for which the nodes and 
connections contribute collectively to its global functionality which includes transport processes \cite{Hughes,MasudaPhysRep2017}, 
synchronization \cite{Arenas2008}, and diffusion \cite{BlanchardBook2011,FractionalBook2019}, among others \cite{VespiBook}.  
The capacity of this structure to perform these functions is measured by a global quantity 
in a determined configuration (state) of the system. (2) The entire system is subjected to stochastic 
damage that reduces the functionality of the links affecting the global activity, and this 
detriment is cumulative.  (3) We compare the global functionality of the 
system with the initial state (with optimal conditions and no damage) and define a 
threshold value for the functionality 
required for the system to operate, i.e., to be alive. 
\\[2mm]
The three features may exist in different complex systems. 
The gradual deterioration of the global functionality resulting from the accumulation of misrepairs in a complex system 
is the subject of the model to be developed and explored in the present section. The model incorporates the 
observed phenomena of self-amplification in the occurrence of misrepairs; i.e., a structure 
that is already altered by misrepairs is more likely to `attract' further misrepairs \cite{WangMi2015b}. We also account for the observation that there are two temporal scales. One is the ``fast'' timescale of functional operation; for example, the temporal evolution of a random walk. The second ``slow'' timescale is where the dynamics of accumulation of misrepairs and aging take place. These assumptions reflect the observed fact that functions in a living organism may take seconds or minutes, whereas aging changes in living beings may take years.
\subsection{Network structure and cumulative damage}
We consider undirected connected networks with $N$ nodes $i=1,\ldots ,N$. The topology of the network is described by an adjacency matrix $\mathbf{A}$ with elements $A_{ij}=A_{ji}=1$ if there is an edge between the nodes $i$ and $j$ and $A_{ij}=0$ otherwise; in particular, $A_{ii}=0$ to avoid lines connecting a node with itself. 
The degree of the node $i$ is the number of its neighbor nodes and given by $k_i=\sum_{l=1}^N A_{il}$. In this structure, we denote the set of nodes as $\mathcal{V}$ and the set of lines (links, edges) $\mathcal{E}$ with elements $(i,j)$. For each pair in $\mathcal{E}$, the corresponding element of the adjacency matrix is non-null. Due to the symmetry of the adjacency matrix $(i,j)$ is equivalent to $(j,i)$. In the following, we denote as  $|\mathcal{E}|$ the total number of different lines in the network.
\\[2mm]
Additionally to the network structure, the global state of the system at time $T=0,1,2,\ldots$ is 
characterized by a $N\times N$ symmetric matrix $\mathbf{\Omega}(T)$ with elements $\Omega_{ij}(T)=\Omega_{ji}(T)\geq 0$ and $\Omega_{ii}(T)=0$, which describe weighted connections between the nodes. The matrix $\mathbf{\Omega}(T)$ contains information of the state of the edges. 
Now, in order to capture in the model the damage impact affecting the complex system, we introduce a variable $T$ as a measure of the number 
of damage hits in the links of the network. $T$ 
can also be conceived as a time measure if we assume constant damage impact rate, i.e., successive damage events occur 
with a constant difference of times $\Delta T=1$. We introduce for each line $(i,j) \in \mathcal{E}$ a random integer variable $h_{ij}(T)$ 
where $h_{ij}(T)-1$ counts the number of 
random faults that exist in this link at time $T$. The values $h_{ij}(T)$ for all 
the lines are numbers that evolve randomly, and a new fault in the link $(i,j)$ appears 
at time $T$ with a probability $\pi_{ij}(T)$ which is given by
\begin{equation}\label{problinks}
\pi_{ij}(T)=\frac{h_{ij}(T-1)}{\sum_{(l,m) \in \mathcal{E}} h_{lm}(T-1)}\qquad (i,j) \in \mathcal{E},
\end{equation}
for $\,T=1,2,\ldots$ with the initial condition $h_{ij}(0)=1$, i.e. no faults exist for all the edges 
at $T=0$ (`birth' of the complex system). 
The relation (\ref{problinks}) indicates the probability for the  event that at time $T$ the number of faults $h_{ij}(T)=h_{ij}(T-1)+1$ are increased by one.
For simplicity in our analysis and to maintain the network undirected, 
we assume $h_{ij}(T)=h_{ji}(T)$ and also $\pi_{ij}(T)=\pi_{ji}(T)$. 
With Eq. (\ref{problinks}) at $T=1$ is randomly generated the first hit (fault) for any selected line $(i,j)$ 
with equal probability. 
The occurrence of the second fault at $T=2$ depends on the previous configuration and so on. An essential feature of
the probabilities in Eq. (\ref{problinks}) is that they produce preferential damage if a link has already 
suffered damage in the past. A link has a higher probability to get a fault with respect to a 
line never being damaged. 
Such preferential random processes have also been explored in different contexts in science; 
see Ref. \cite{NetworkScienceBook2016}. 
In appendix Section \ref{Section_appendixA} 
we present a detailed analysis of how Eq. (\ref{problinks}) 
produces a hierarchical distribution of damage in the lines. 
\\[2mm]
The choice of generating law of faults
(\ref{problinks}) is based on the observation of aging changes in living beings. Development of aging 
changes such as age spots is self-amplifying and inhomogeneous. Age spots develop in this way 
because misrepaired structures have increased damage sensitivity
and reduced reparation-efficiency \cite{WangMi2015,WangMi2015b}.
\\[3mm]
We aim to describe how the structure reacts to the damage hits occurring randomly to the lines. 
We describe the effects of the damage by using the information in 
the matrix of weights $\mathbf{\Omega}(T)$. 
In terms of the values $h_{ij}(T)$, the matrix $\mathbf{\Omega}(T)$ defines 
the global state of the network at time $T$ through the elements
\begin{equation}\label{OmegaijT}
\Omega_{ij}(T)=(h_{ij}(T))^{-\alpha} A_{ij},
\end{equation}
where $\alpha\geq 0$ is a real valued parameter that quantifies 
the effect of the damage in each link. We call $\alpha$ the {\it misrepair parameter} 
since it describes the capacity of the system to repair a damage in the links: 
In the limit $\alpha\to 0$ the system responds with perfect 
reparation with $\Omega_{ij}(T) \to A_{ij}$ as in a perfect undamaged structure, and 
the effect of the stochastically generated faults is null. On the other hand, 
in the limit $\alpha\to\infty$, a hit in a line is equivalent to its removal from the network. This limit 
corresponds to a complex system without repair capacity. The fault accumulation 
dynamics described by Eq. (\ref{problinks}) together with the `misrepair equation' (\ref{OmegaijT}) is therefore able to mimic the phenomena related to aging processes observed in living beings \cite{WangMiWun2009,Kirkwood2005}.
\begin{figure*}[!t]
\begin{center}
\includegraphics*[width=1.0\textwidth]{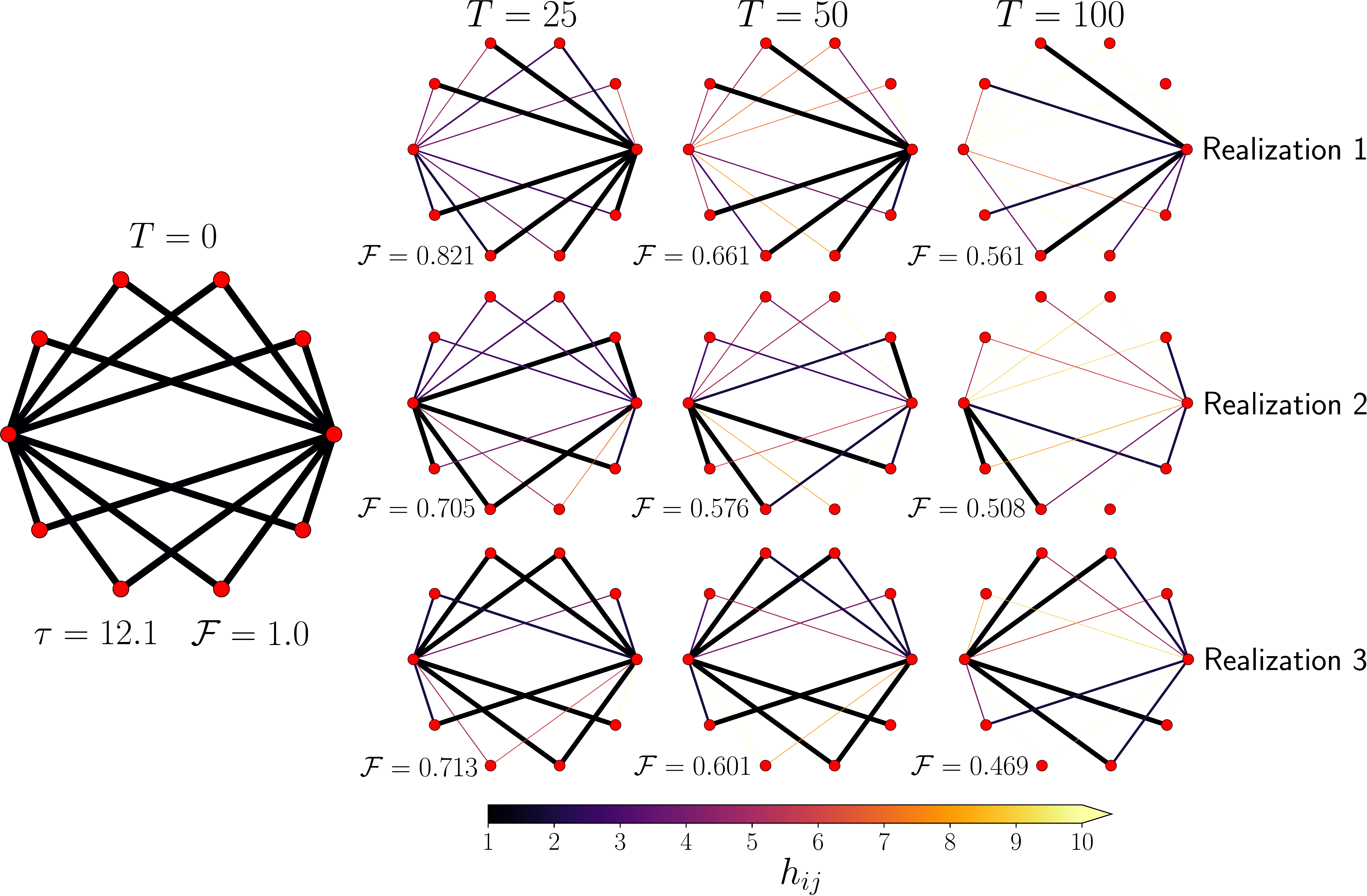}
\end{center}
\vspace{-5mm}
\caption{\label{Fig_1} (Color online)  Monte Carlo simulation of the reduction of functionality in networks with random faults in the lines. We generate a random hit (fault) in the lines at times $T=1,2,\ldots$ and in each line we have a value $h_{ij}$ that depends on $T$. At the time $T$, the value $h_{ij}-1$ gives the number of random hits that the line $(i,j)$ has suffered (considering the initial values $h_{ij}=1$ at $T=0$). The probability to have a new fault in one of the lines is determined by Eq. (\ref{problinks}). By using this algorithm,  we implement Monte Carlo simulations to generate random faults in the network described at time $T=0$. The values in the color bar indicate $h_{ij}$ for all the lines, and the widths represent the values $\Omega_{ij}(T)$ that describe their capacity of transport reduced with the increment of $h_{ij}$; in all the realizations we use the misrepair parameter $\alpha=2$ in Eq. (\ref{OmegaijT}). We depict the results of three independent realizations of this process and 
we present the configurations of the system at times $T=0,\, 25,\, 50,\, 100$ along with the global value $\mathcal{F}$ in Eq. (\ref{F_ratioT}) that determines the global functionality of the structure being $\mathcal{F}=1.0$ for the initial case without damage [in this initial configuration the value $\tau=12.1$ is calculated by using Eq. (\ref{tauglobal})]. The functionality  $\mathcal{F}$ evolves with time and gradually is reduced with the faults in the lines.}
\end{figure*} 
\subsection{Transport and global functionality}
In the above described system the random walk in the network with discrete steps at times $t=0,1,\ldots$
takes place  
at a significantly smaller timescale as the characteristic time interval $\Delta T$ of the damage impacts.
For each configuration of the network weights ${\mathbf \Omega}(T)$ the walker eventually visits all nodes in the structure that changes very slowly with time $T$.
In a determined configuration at time $T$, the transition probability of the random walker 
to pass from node $i$ to node $j$ is given by
\begin{equation}\label{transitionPij}
w_{i\to j}(T)=\frac{\Omega_{ij}(T)}{\sum_{\ell=1}^N\Omega_{i\ell}(T)}.
\end{equation}
We assume that the random walker is Markovian and governed by a master equation that 
describes the temporal evolution as well as the exploration of the network in the configuration at time $T$ 
(see the appendix in Section \ref{Section_appendixB} for details).
\\[2mm]
In terms of the random walker defined in Eq. (\ref{transitionPij}), 
we can measure the global transport capacity of the structure by using the characteristic time $\tau(T)$
which is defined as \cite{RiascosMateos2012,RiascosMichelitsch2017_gL} 
\begin{equation}\label{tauglobal}
\tau(T)\equiv\frac{1}{N}\sum_{i=1}^{N}\tau_i(T) \,  ,
\end{equation}
with  \cite{NohRieger2004,RiascosMateos2012,RiascosMichelitsch2017_gL}
\begin{figure*}[!t]
\begin{center}
\includegraphics*[width=0.91\textwidth]{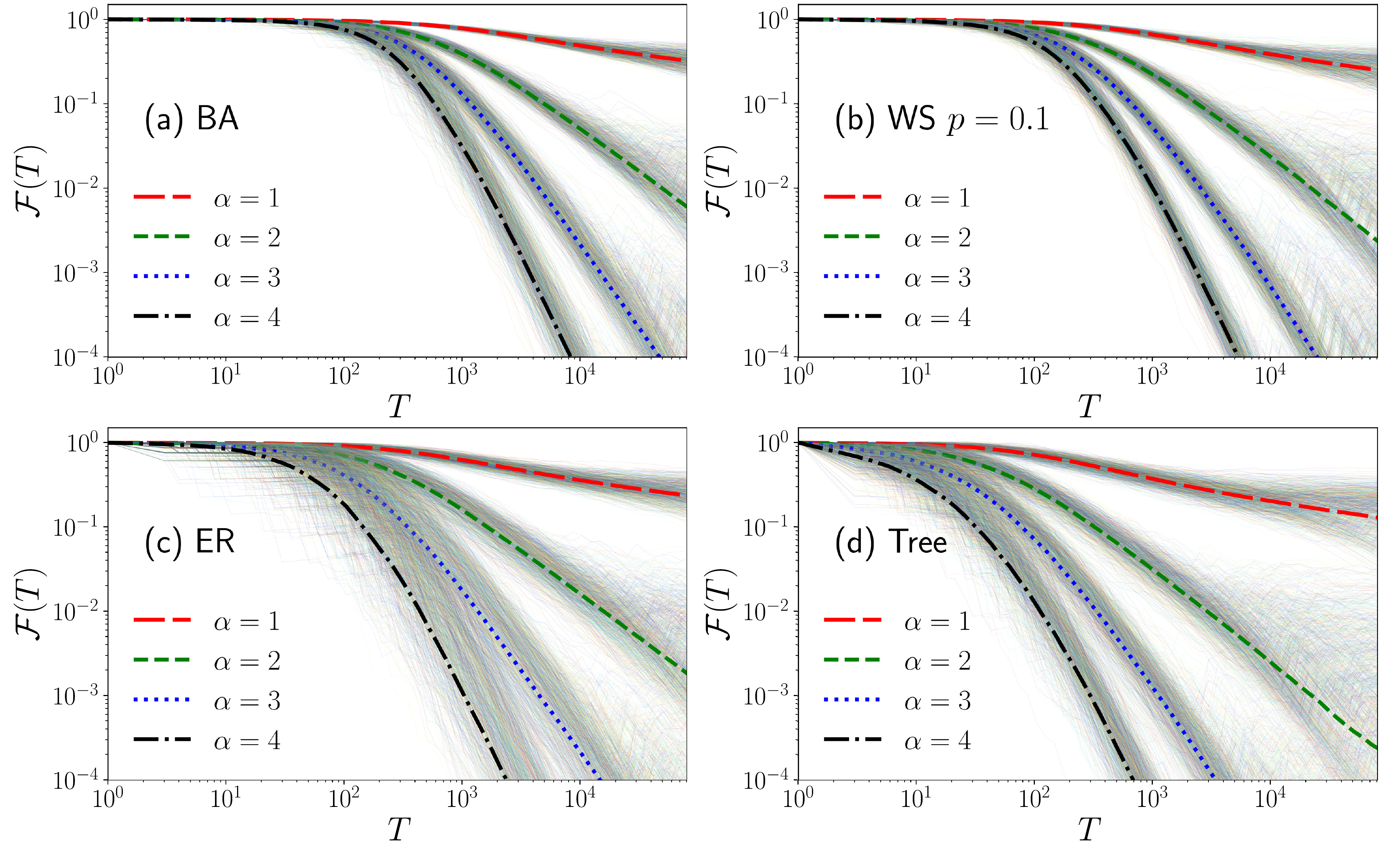}
\end{center}
\vspace{-5mm}
\caption{\label{Fig_2} (Color online) Evolution of the functionality $\mathcal{F}(T)$  
in connected networks with random faults in the lines. 
We implement the algorithm described in Eqs. (\ref{problinks}) and (\ref{OmegaijT}) 
to simulate the reduction of the capacity of transport in different types of networks with $N=100$ nodes: (a) Barab\'asi-Albert (BA) network, (b) 
Watts-Strogatz (WS) network with rewiring probability $p=0.1$, (c)  Erd\"{o}s-R\'enyi (ER) network at the percolation limit $p=\log(N)/N$, and (d) a tree.  We depict the numerical results of the functionality $\mathcal{F}(T)$ given by Eq. (\ref{F_ratioT}). We simulate $1000$ realizations of this process for the misrepair parameters $\alpha=1$, $\alpha=2$, $\alpha=3$, $\alpha=4$ and show with wide dashed lines the results obtained for the average over realizations $\langle\mathcal{F}(t)\rangle$  for each $\alpha$. With different thin lines we depict the values of $\mathcal{F}(t)$ obtained in each realization.}
\end{figure*} 
\begin{equation}\label{TauiSpect}
    \tau_i(T)=\sum_{l=2}^N\frac{1}{1-\lambda_l(T)}\frac{\left\langle i|\phi_l(T)\right\rangle \left\langle\bar{\phi}_l(T)|i\right\rangle}{\left\langle i|\phi_1(T)\right\rangle \left\langle\bar{\phi}_1(T)|i\right\rangle}\, ,
\end{equation}
where $\{\lambda_i(T)\}_{i=1}^N$ are the eigenvalues of the transition matrix $\mathbf{W}(T)$ 
with elements given by Eq. (\ref{transitionPij}) [we always denote $\lambda_1(T)=1$]. 
In the same way, we denote as $\left|\phi_i(T)\right\rangle$ and $ \left\langle\bar{\phi}_i(T)\right|$ its 
respective right- and left eigenvectors ($i=1,2,\ldots, N$). 
\\[2mm]
Therefore, for each global configuration at time $T$ we have the transition probability matrix
$\mathbf{W}(T)$ and by using its eigenvalues and eigenvectors we determine the characteristic 
time in Eq. (\ref{tauglobal}) that quantifies the capacity of transport of the system. 
To indicate explicitly that this characteristic time describes the transport in the global 
configuration of the system at time $T$, we denote the quantity in Eq. (\ref{tauglobal}) as $\tau(T)$. 
This quantity gives an estimate of the average number of steps the walker needs to reach any site 
of the network (see appendix in Section \ref{Section_appendixB})
and is therefore an important measure to characterize the capacity of a random walker 
to visit the nodes of a network at a determined configuration at time $T$ \cite{NohRieger2004,RiascosMateos2012}. 
\\[2mm]
In particular, for the initial configuration at $T=0$, the elements of the matrix of weights satisfy $\Omega_{ij}(0)=A_{ij}$; therefore, in this case, the random walker follows the transition matrix $\mathbf{W}(0)$ for a normal random walk in a network with $w_{i\to j}(0)$ given by \cite{NohRieger2004}
\begin{equation}
w_{i\to j}(0)=\frac{A_{ij}}{k_i}.
\end{equation}
For this particular random walk strategy, we denote the global time $\tau(T)$ in Eq. (\ref{tauglobal}) as
\begin{equation}\label{tau0def}
\tau_0\equiv\tau(0).
\end{equation}
Finally, we define the `functionality' $\mathcal{F}(T)$ that quantifies the global transport capacity of the 
network at time $T$ as
\begin{equation}\label{F_ratioT}
\mathcal{F}(T)\equiv\frac{\tau_0}{\tau(T)}.
\end{equation}
The functionality $\mathcal{F}(T)$ characterizes globally the effect of the damage suffered 
by the whole structure and how evolves the capacity of a 
random walker to explore the network. The smaller the $\tau(T)$ (i.e., the higher the transport capacity), 
the higher the functionality. Since the time $\tau(T) \geq \tau_0$ in the damaged structure is 
greater than in the undamaged structure we have $\mathcal{F}(T) \leq 1$ 
(equality holds only in the undamaged state).
\\[2mm]
An intuitive interpretation of the degradation of transport capacity in the network is the following. With increasing 
time $T$ a large number of very weakly connected edges $(u,v)$ containing large fault 
numbers $h_{uv}\gg 1$ are generated with 
$\Omega_{uv}(T) \sim (h_{uv}(T))^{-\alpha} \to 0$. As a consequence
an increasing number of weakly connected or quasidisconnected regions emerges with 
low transition probabilities between
these weakly connected parts. The walker hence remains trapped for many steps in these regions, 
which considerably increases the global characteristic time the walker needs 
to reach any node of Eq. (\ref{tauglobal}). 
This is also reflected by the fact that
for each 
of these weakly connected regions an eigenvalue $\lambda \to 1$ close to one emerges in the
transition matrix which generates singular behavior in the sums of Eq. (\ref{TauiSpect}).
\\[3mm]
In Fig. \ref{Fig_1}, we illustrate the concepts introduced in this section for a network with $N=10$ nodes. 
We generate random hits in the lines of this network at each time $T=1,2,\ldots,100$. The probability to generate a fault at time $T$ in the line $(i,j)$ is proportional to the previous configuration given by $h_{ij}(T-1)$ in Eq. (\ref{problinks}). 
By using Monte Carlo simulations, we recreate the damage in three different realizations. 
The values of $h_{ij}$ in the lines are represented with colors 
codified in the color bar,
whereas the respective widths reflect 
their capacity of transport given by $\Omega_{ij}(T)$ in Eq. (\ref{OmegaijT}) 
with the parameter $\alpha=2$. 
For each configuration of the system, we calculate the transition probability matrix $\mathbf{W}(T)$ 
in Eq. (\ref{transitionPij}) by using Eq. (\ref{OmegaijT}) to obtain numerically the 
time $\tau(T)$ of Eq. (\ref{tauglobal}). 
Finally, the relation between  $\tau(0)$ and $\tau(T)$ allows 
us to establish the global functionality $\mathcal{F}(T)$ in Eq. (\ref{F_ratioT}), 
a quantity that evolves with $T$. All this is shown in Fig. \ref{Fig_1} 
for different realizations, where we depict the configurations and functionality at $T=0,\,25,\, 50,\, 100$.  In the 
Supplemental Material, we present videos with the complete simulations \footnote{See supplemental material for videos with Monte Carlo simulations of this algorithm for  $\alpha=2$ and $T=0,1,\ldots,100$. Three different realizations are presented in the videos video1.avi, video2.avi, video3.avi, respectively.}. In general, we observe that the configurations at $T=100$ differ in each realization as well as their global functionality.
\section{Global functionality and lifespan}
\begin{figure*}[!t]
\begin{center}
\includegraphics*[width=1.0\textwidth]{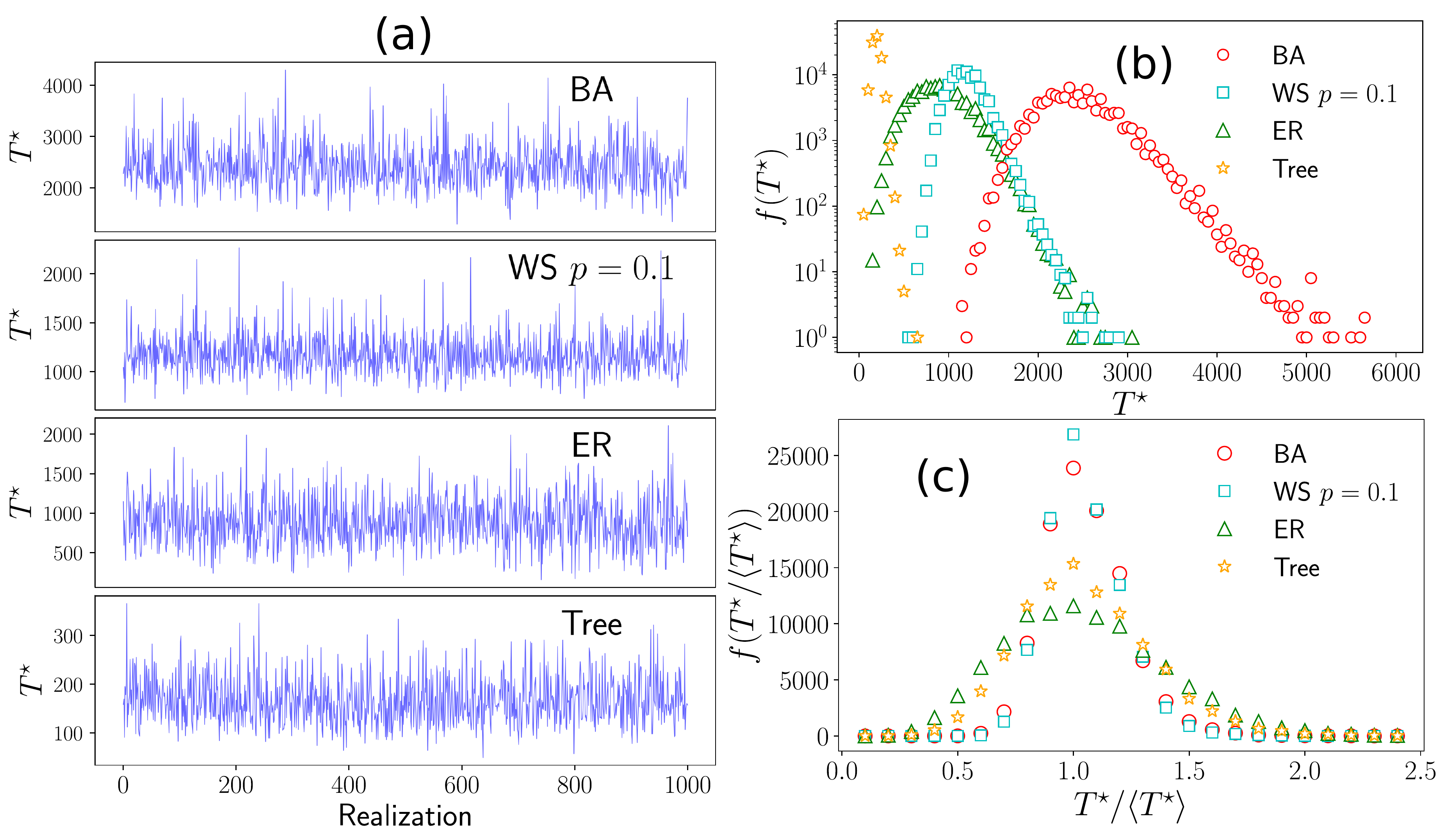}
\end{center}
\vspace{-5mm}
\caption{\label{Fig_3}  (Color online) Statistical analysis of the lifetime of networks under stochastic damage and misrepair. We consider stochastic faults in the lines, the misrepair parameter $\alpha=2$ in Eq. (\ref{OmegaijT}) and, the threshold value $\mathcal{F}^\star=0.2$ to define the limit of functionality in Eq. (\ref{Tstar_life}). (a) Lifetime $T^{\star}$ of $1000$ realizations of the process in several types of networks with $N=100$ nodes: Barab\'asi-Albert (BA), Watts-Strogatz (WS) with rewiring probability $p=0.1$, Erd\"{o}s-R\'enyi (ER), and a tree. (b) Frequencies $f(T^{\star})$ of the times $T^{\star}$ obtained in $10^5$ realizations. In panel (c) we express the frequencies of the times $T^{\star}$ in terms of their average over realizations $\langle T^{\star}\rangle$.}
\end{figure*} 
We defined in Section \ref{SectionModel} a model to analyze the effects of the damage in the global transport capacity of a network; in the following part, we explore the functionality  $\mathcal{F}(T)$ in the context of aging due to cumulative damage and misrepair in different types of networks. The reduction of this functionality allows us to define characteristic times associated with the lifespan and the aging in each structure.
\\[2mm]
In Fig. \ref{Fig_2}, we present the results of Monte Carlo simulation for the algorithm of damage introduced in Eqs. (\ref{problinks})-(\ref{F_ratioT}) for different types of networks. Our simulations are similar to the examples presented in Fig. \ref{Fig_1}. We analyze the evolution of the system subjected to stochastic damage in the lines in a Barab\'asi-Albert network generated with a preferential attachment algorithm \cite{BarabasiAlbert1999}, a Watts-Strogatz network with rewiring probability $p=0.1$ \cite{WattsStrogatz1998}, an Erd\"{o}s-R\'enyi network at the percolation limit \cite{ErdosRenyi1959} and a tree. We analyze the values of $\mathcal{F}(T)$ as a function of $T$ for these structures in different realizations. For the values $\alpha=1, \,2,\, 3,\, 4$, we see that the measure of the global time $\tau(T)$ in Eq. (\ref{tauglobal}) differs slightly from the previous value $\tau(T-1)$, i.e. $|\tau(T)-\tau(T-1)|\ll\tau_0$. In addition $\tau(T)-\tau(T-1)$ may be positive or negative due to the fact 
that in particular states, the reduction of the global functionality of a line could produce a small increment of 
the functionality. However, in general, the most common effect is the damage of the structure, 
and therefore, we see for $\alpha>0$ that $\mathcal{F}(T)$ starts in $\mathcal{F}(0)=1$ and gradually is reduced with the increase of $T$ in each realization. In Fig. \ref{Fig_2}, we also present the average over $1000$ realizations and from the small deviations observed we can infer that the ensemble average $\langle \mathcal{F}(T)\rangle$ is a good description of the aging in the system, i.e., the global reduction of the functionality. 
It is worthwhile to mention that due to the normalization term in the transition probabilities 
in Eq. (\ref{transitionPij}); once all the lines have suffered at least one hit, the transition probabilities  rescale maintaining the same proportion of damage but increasing the values of $\mathcal{F}(T)$  [we can see this in some realizations in Fig. \ref{Fig_2}(b) in the 
variations of $\mathcal{F}(T)$ for $T>10^4$ and $\alpha=2$]. This effect occurs at large times $T$ for structures with a large number of lines; in this case,  the failures concentrate in particular connections,  leaving intact other links of the system.
\\[2mm]
The evolution of the systems under damage explored before opens the question: 
When the system is unable to perform correctly the function assigned? 
To describe this failure effect we assume that there exists a minimal functionality $\mathcal{F}^\star$ 
required for the system to operate. Once defined this threshold value, there is a lifetime $T^{\star}$ that satisfies
\begin{equation}\label{Tstar_life}
T^{\star}=\min\{T=1,2,\ldots| \mathcal{F}(T)<\mathcal{F}^\star\}.
\end{equation}
We use this definition to identify the first time for which the functionality of the system is 
below the threshold value $\mathcal{F}^ \star$. Therefore, for times $T\leq T^\star$ 
the system is ``alive'' since it can perform the operation assigned. For $T> T^\star$ the system ``dies.''
\\[2mm]
For a given value $ \mathcal{F}^\star$, the lifetime $T^\star$ varies in each of the realizations; however, 
the ensemble average $\left\langle T^\star \right\rangle$ is a characteristic of each network. 
We can conceive $\left\langle T^\star \right\rangle$ as life-expectancy of the system.
In Fig. \ref{Fig_3} we analyze statistically the times $T^\star$ for the networks with $N=100$ nodes in Fig. \ref{Fig_2}, and we consider $\mathcal{F}^\star=0.2$ and $\alpha=2$. In Fig. \ref{Fig_3}(a) we show how $T^\star$ varies in the realizations, and in Fig. \ref{Fig_3}(b) we explore the frequencies of the lifetime in the results obtained with $10^5$ realizations of the system. We see that the Barab\'asi-Albert network is the most resilient structure with the highest values of lifespan. The Watts-Strogatz network also presents high values of the lifetime.  In comparison with these complex networks, the most frequent values of $T^\star$ for the  Erd\"{o}s-R\'enyi network at the percolation limit 
and the tree reveal that the lifetimes in these two systems are lower, the tree is the structure with the lowest values of $T^\star$. This last result makes sense since in a tree, the complete removal of a line immediately disconnects the structure. On the other hand, complex networks like the Barab\'asi-Albert 
network have several redundant paths connecting the nodes making them more resistant to damage. 
In Fig. \ref{Fig_3}(c), we represent the lifetimes $T^\star$ as a fraction of the 
ensemble average $\left\langle T^\star \right\rangle$ over the realizations. We analyze the frequencies of the 
values  $T^\star/\left\langle T^\star \right\rangle$ to see how they distribute 
around the ensemble average $\left\langle T^\star \right\rangle$. In this way, we can infer 
that the ensemble average of $T^{\star}$ 
is a good measure for the life-expectancy since it is 
close to the most probable lifetime as shown by the peak around $T^\star/\left\langle T^\star \right\rangle=1$. 
We see that values with 
$ T^\star \gg \left\langle T^\star \right\rangle $ or $ T^\star \ll \left\langle T^\star \right\rangle $ 
appear with very low probability. Finally, we also observe similarities in the frequencies calculated for the Barab\'asi-Albert and the Watts-Strogatz network; the Erd\"{o}s-R\'enyi network and the tree also share similar characteristics.
\\[2mm]
\begin{table}[!t]
\centering
\caption{\label{Table1} Description of the networks with $N=100$ nodes explored in Fig. \ref{Fig_3} (presented in the first four rows) and Fig. \ref{Fig_4} [for the Watts-Strogatz (WS) networks with rewiring probability $p=0,\, 0.1,\ldots,\, 1.0$]. $|\mathcal{E}|$ is the number of edges, $\bar{k}$ denotes the average degree, $\bar{d}$ is the average distance between nodes, and $\tau_0$ is calculated for the initial structure with no damage and describes a normal random walker in the network. Also, we obtain the lifetimes $T^{\star}$ for $10^5$ realizations of the process with misrepair parameter $\alpha=2$ and $\mathcal{F}^{\star}=0.2$; we present the ensemble average $\left\langle T^\star \right\rangle $ with the respective standard deviation $\sigma_{T^\star}$.}
\vspace{3mm}
\begin{tabular}{l c c c c c c}
\hline
\hline    
{\bf Network} & $|\mathcal{E}|$ & $\bar{k}$ & $\bar{d}$ & $\tau_0$ & $\langle T^\star\rangle$ & $\sigma_{T^\star}$\\
\hline
Barab\'asi-Albert          & 294 & 5.88  & 2.51   & 163.58 & $2414.4$  & $423.4$ \\
Watts-Strogatz             & 200 & 4     & 5.27   & 256.76 & $1171.6$  & $186.9$\\
Erd\"{o}s-R\'enyi          & 214 & 4.28  & 3.33   & 196.18 & $878.8$  & $302.1$\\
Tree                       & 99 & 1.98   & 7.73   & 758.17 & $168.2$   & $48.0$\\
\hline 
WS $p=0.0$                & 200 & 4     & 12.88       & 702.22  & 1087.9   &  184.1 \\
WS $p=0.1$                & 200 & 4     & 4.79        & 242.44  & 1161.4   &  186.3 \\
WS $p=0.2$                & 200 & 4     & 4.00        & 187.81  & 1154.8   &  198.9 \\
WS $p=0.3$                & 200 & 4     & 3.91        & 183.19  & 1130.3   &  198.3 \\
WS $p=0.4$                & 200 & 4     & 3.62        & 181.68  & 977.7    &  261.8 \\
WS $p=0.5$                & 200 & 4     & 3.56        & 175.63  & 1038.6   &  225.6 \\
WS $p=0.6$                & 200 & 4     & 3.56        & 178.56  & 981.8    &  240.7 \\
WS $p=0.7$                & 200 & 4     & 3.47        & 181.54  & 970.3    &  226.4 \\
WS $p=0.8$                & 200 & 4     & 3.54        & 187.43  & 904.0    &  232.2 \\
WS $p=0.9$                & 200 & 4     & 3.47        & 185.51  & 916.5    &  231.2 \\
WS $p=1.0$                & 200 & 4     & 3.44        & 195.32  & 831.7    &  230.3 \\
\hline 
\hline  
\end{tabular}
\end{table}
\begin{figure}[!t]
\begin{center}
\includegraphics*[width=0.49\textwidth]{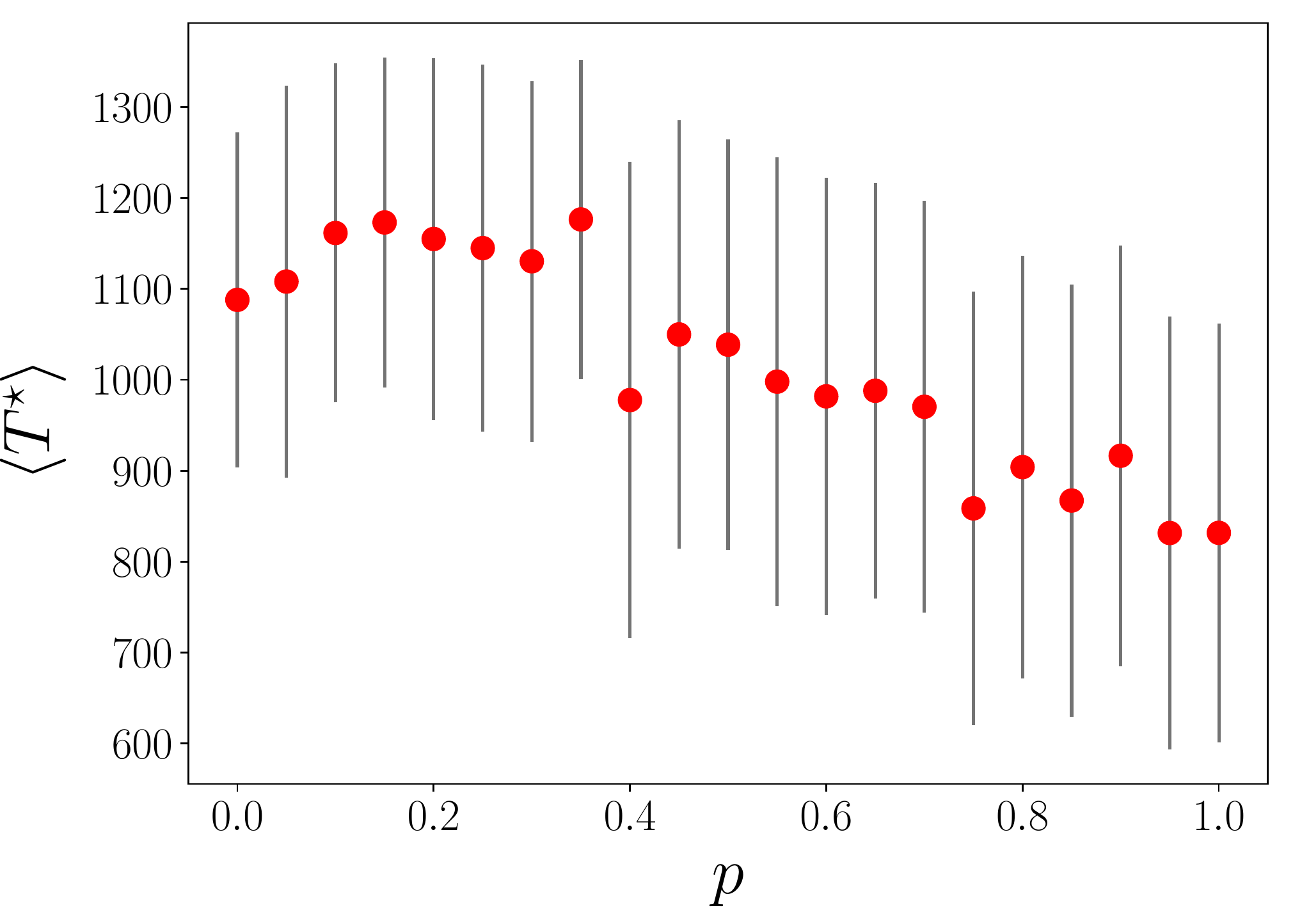}
\end{center}
\vspace{-6mm}
\caption{\label{Fig_4} (Color online) Ensemble average $\left\langle T^\star \right\rangle$ of the times $T^{\star}$ in connected Watts-Strogatz networks with rewiring probability $p$. The results for each network were calculated from $10^5$ realizations with misrepair parameter $\alpha=2$ and considering $\mathcal{F}^{\star}=0.2$. The error bars were obtained from the standard deviation of the data $\sigma_{T^\star}$.}
\end{figure} 
Furthermore, from the results in Fig. \ref{Fig_3} and the model introduced, we can infer that one important feature in the network topology that influences the value of $T^\star$ is the number of lines $|\mathcal{E}|$  in the network. Higher values of $|\mathcal{E}|$ could make a structure more resilient to damage. However, how these lines are connected is of utmost importance in the lifetime of the system. For example, when we consider the global transport, in some configurations, the complete removal of a single line could disconnect a part of the network making null the respective functionality independently of the number of lines. In Table \ref{Table1}, we present different quantities that characterize the structure of the networks explored in Figs. \ref{Fig_2} and \ref{Fig_3}. 
We include the number of lines $|\mathcal{E}| =\frac{1}{2}\sum_{i=1}^N k_i$, the average degree $\bar{k}\equiv \frac{1}{N}\sum_{i=1}^N k_i$, and 
the average distance $\bar{d}$ between nodes
\begin{equation*}
\bar{d}\equiv \frac{1}{N(N-1)}\sum_{i=1}^N\sum_{j=1}^N d_{ij},
\end{equation*}
where $d_{ij}$ is the number of lines in the shortest path in the network connecting the nodes $i$ and $j$. We also present the time $\tau_0$ in Eq. (\ref{tau0def}) that gives an estimate of the average number of steps needed by a normal random walker to reach any node in the network, in this way this is a measure of the capacity of the structure to connect their nodes with higher values for networks that are difficult to explore and a minimum value $\tau_0=(N-1)^2/N$ in fully connected networks  
\cite{FractionalBook2019}. For each network, we include the ensemble average lifetime $\langle T^\star \rangle$ and 
the respective  standard deviation $\sigma_{T^\star}$ that measures the spread of the times $T^\star$ for the $10^5$ 
realizations in the Monte Carlo simulations with $\mathcal{F}^{\star}=0.2$ analyzed in Fig. \ref{Fig_3}. 
In these results, we see the connection between $\bar{d}$ and $\tau_0$ 
for which higher values of the average distance require longer times of exploration $\tau_0$. 
However, from the different measures, it is unclear the relation of $\langle T^{\star} \rangle$ with the 
quantities presented to describe globally each network. Nevertheless, intuitively, we can infer that the 
complexity of the structure 
plays an important role since the average times $\langle T^\star\rangle$ are higher for networks of the Barab\'asi-Albert and Watts-Strogatz type and are smaller in networks with a simpler structure like the tree and the Erd\"{o}s-R\'enyi network at the percolation limit.
\\[2mm]
Now, in order to have more evidence about the relation between $T^{\star}$ and the complexity of the network, we will analyze random networks conserving the same number of lines  $|\mathcal{E}|$. In Fig. \ref{Fig_4} we  explore the effects of aging in connected networks generated with the Watts-Strogatz algorithm for different values of the rewiring probability $0\leq p\leq 1$ \cite{WattsStrogatz1998}. In the  Watts-Strogatz algorithm, for $p=0$, the network is regular with degree $k=4$ and is a ring with additional links to connect each node with its four nearest nodes on the ring. Then, the extreme of a $p$ fraction of links is relocated randomly producing connections with distant nodes \cite{WattsStrogatz1998}. For $p$ small the complexity of the network increases with respect to the regular structure with $p=0$ due to the rewiring. However, for values of $p$ close to 1, the high random rewiring produces a less complex disordered structure that in the limit $p\to 1$ is equivalent to an Erd\"{o}s-R\'enyi network. In Fig. \ref{Fig_4}, this behavior is 
observed for the times $\left\langle T^\star \right\rangle$ for Watts-Strogatz networks with rewiring probability $p$. In particular, for $p=0$,  $\left\langle T^\star \right\rangle=1087.9$ and the values of $\left\langle T^\star \right\rangle$ are higher for $0<p<0.4$. For  $0.5<p\leq 1.0$ the values of $\left\langle T^\star \right\rangle$ decrease and are lower than the lifetime of the network with $p=0$. The minimum values of $\left\langle T^\star \right\rangle$ are found for $p=0.95$ and $p=1.0$. In Table \ref{Table1}, we present the numerical values of the ensemble average of the lifetime for the networks with $p=0, 0.1,\ldots, 0.9, 1.0$ and some characteristic values that describe each network. With these results, we reaffirm that, although the number of lines in the networks is the same, it is hard to establish a global characteristic of the network that could describe the average value of the system's lifetime. 
We suggest that the value $\langle T^{\star}\rangle$ itself can be used as a new measure 
that contains important quantitative information about the complexity of a network 
and its resistance to damage in the connections.
\\[2mm]
\begin{figure}[!t]
\begin{center}
\includegraphics*[width=0.49\textwidth]{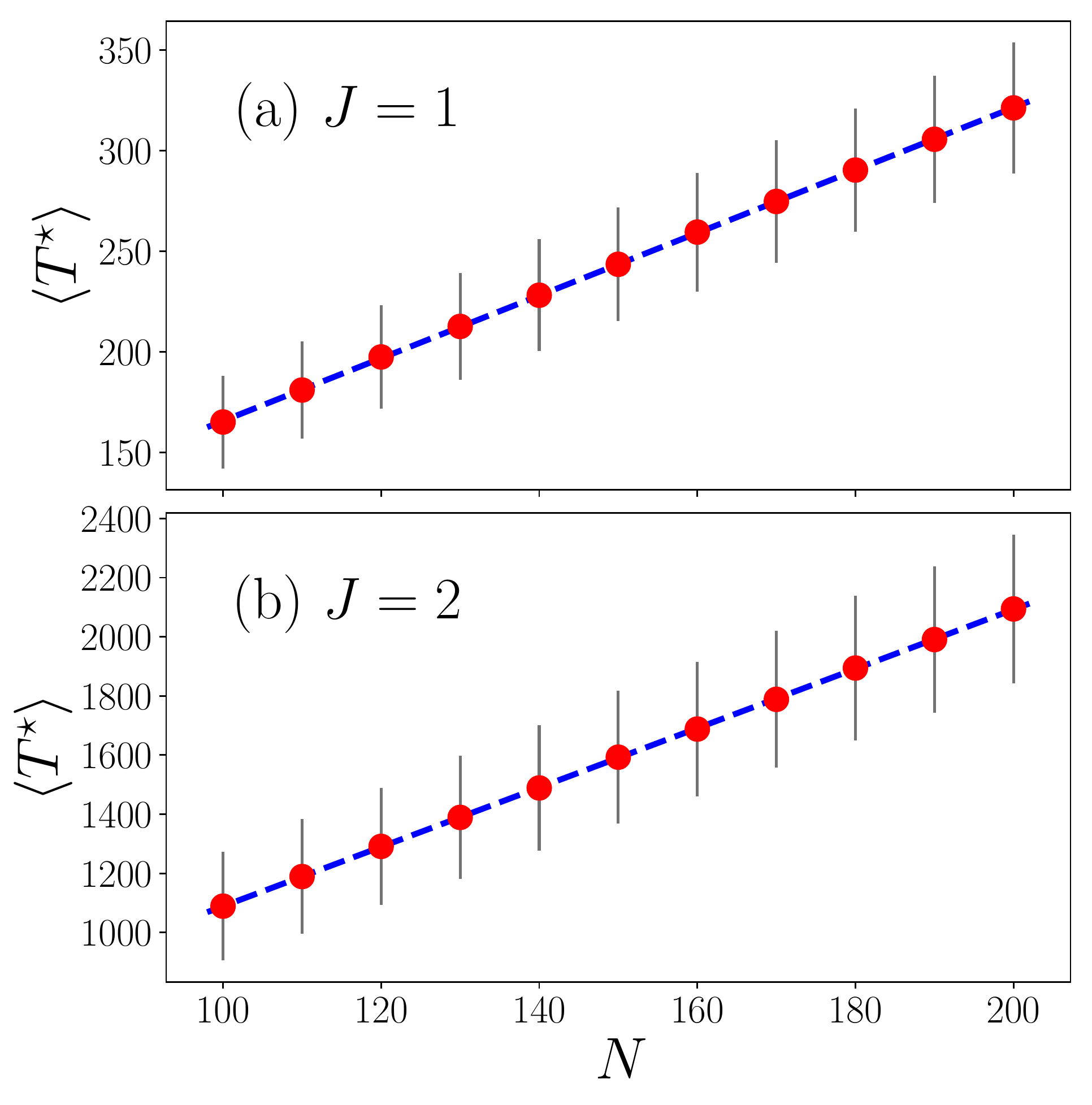}
\end{center}
\vspace{-6mm}
\caption{\label{Fig_5} (Color online) Ensemble average $\left\langle T^\star \right\rangle$ as a function of $N$ for interacting cycles with (a) $J=1$ and (b) $J=2$. The results for each network were calculated from $10^4$ realizations with a misrepair parameter $\alpha=2$ and a functionality threshold $\mathcal{F}^{\star}=0.2$. The error bars were obtained with the standard deviation of the data $\sigma_{T^\star}$. Dashed lines represent the linear fit $\left\langle T^\star \right\rangle=a+b\,N$. For the ring with $J=1$, we obtain the values $a=10.069$, $b=1.557$ with a correlation coefficient $r= 0.999960$ and, we have $a=85.654$, $b=10.0324$, $r=0.999964$ for $J=2$.}
\end{figure} 
\begin{figure*}[!t]
\begin{center}
\includegraphics*[width=1.0\textwidth]{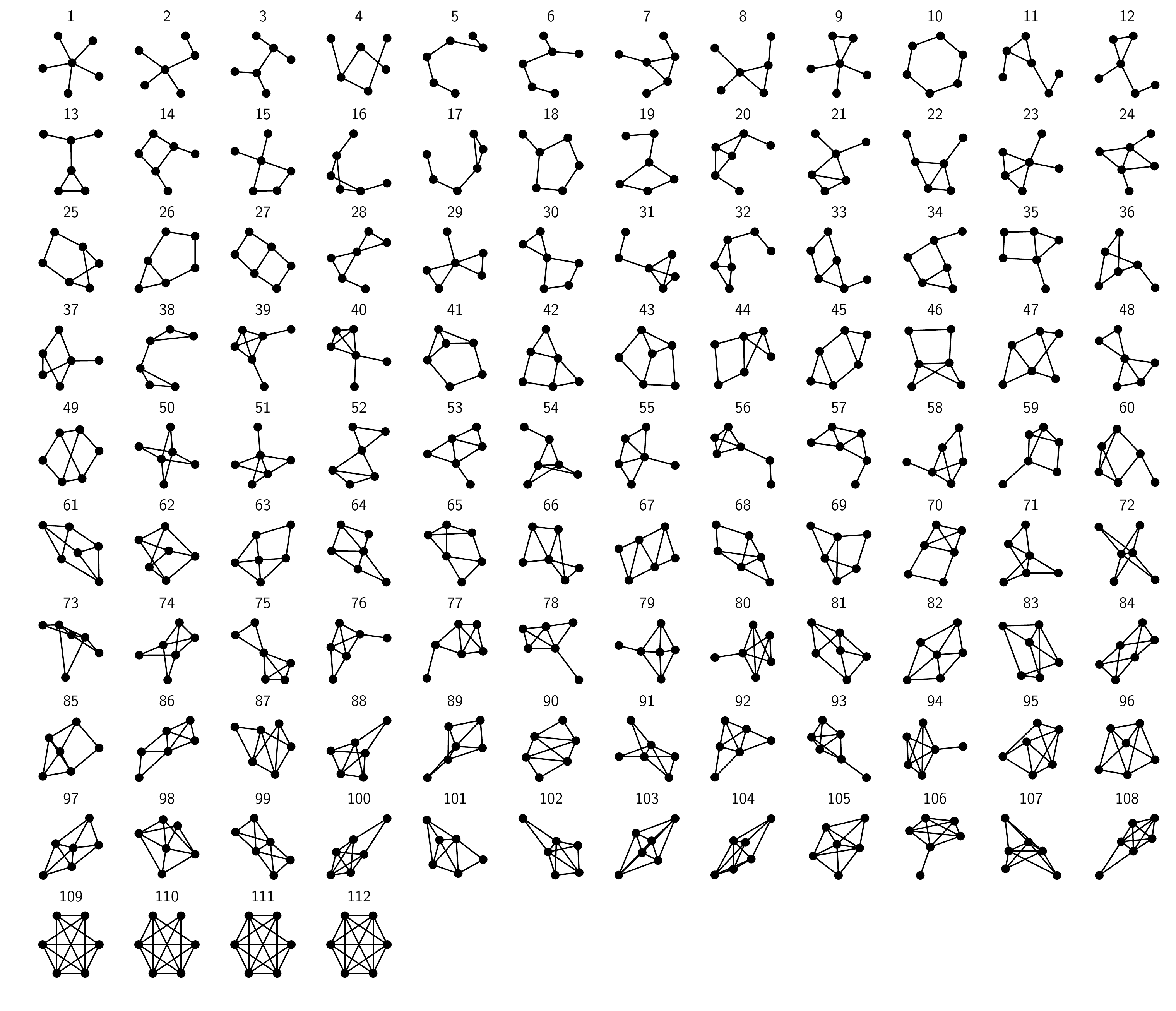}
\end{center}
\vspace{-5mm}
\caption{\label{Fig_6} Nonisomorphic connected graphs with $N=6$ nodes. We sort the structures considering the average of the times $T^\star$ from $10^5$ realizations of the cumulative damage process with the misrepair parameter $\alpha=2$ and threshold functionality $\mathcal{F}^{\star}=0.75$. The average times $\left \langle T^{\star} \right \rangle$ and the respective $\sigma_{T^\star}$ for each configuration are presented in Fig. \ref{Fig_7}.  The catalog of graphs was obtained from Ref. \cite{ConnectedGraphs}.}
\end{figure*} 
To understand the effect of the size of the network, we analyze a particular class of periodic structures called {\it interacting cycles} \cite{VanMieghem,RiascosMateosFD2015,FractionalBook2019}. In these structures, initially, $N$ nodes form a ring. Then, each node is connected to its $J$ left and $J$ right nearest nodes; $2J$ is the degree of the resulting network. The value $J$ is the interaction parameter and all the two nodes whose distance in the initial ring is smaller than or equal to $J$ are connected by additional bonds \cite{RiascosMateosFD2015}. In interacting cycles $|\mathcal{E}|=JN$ is the total number of lines. In particular, $J=1$ defines a ring and $J=2$ determines the initial regular network in the Watts-Strogatz model before rewiring. 
\\[2mm]
In regular networks without damage, the value $\tau(0)$ can be deduced analytically and is given by Kemeny's constant $\tau(0)=\sum_{l=2}^N \frac{1}{1-\lambda_{l}(0)}$ \cite{FractionalBook2019}. By using the eigenvalues $\lambda_{l}(0)$ of the transition matrix  that defines a normal random walker on interacting cycles \cite{VanMieghem,FractionalBook2019}, we have
\begin{equation}\label{TkemenyNN}
\tau(0)=
\sum_{l=2}^N \frac{2J}{2J+1-\frac{\sin\left[\phi_l(2J+1)\right]}{\sin \phi_l}}\, 
\end{equation}
where $\phi_m=\pi(N-m+1)/N$.
\\[2mm]
The introduction of damage in the links at times $T=1,2,\ldots$ breaks the symmetry of the 
structure and gradually reduces its functionality.  We apply our algorithm for the analysis of aging to the study of 
two types of interacting cycles with $J=1$ and $J=2$ and sizes $N=100,110,\ldots,190,200$. In Fig. \ref{Fig_5}, 
we show the numerical results for the average lifespan $\langle T ^\star\rangle$ obtained with  $10^4$ 
realizations. For the different networks explored, our findings reveal that $\langle T ^\star\rangle$ is 
proportional to the size of the network $N$.  In general, we can see that the ring with $J=1$ is more 
vulnerable to damage than the network with $J=2$ for which the existence of more 
lines makes each structure redundant and capable of resisting the damage 
(even the complete removal of some lines does not disconnect the whole structure). The linear 
relation $\left\langle T^\star \right\rangle=a+b\,N$ observed in Fig. \ref{Fig_5} reaffirms 
our previous finding that an important factor in aging is the complexity of the structure. 
This example shows how the variation of $J$ changes the complexity of the network. 
We can say that networks with the same $J$ but different sizes $N$ have the same complexity; 
thus the life expectancy $\langle T ^\star\rangle$ changes for different
line numbers $|\mathcal{E}|=JN$. However, in Fig. \ref{Fig_5} the life expectancy turns out to be much more 
sensitive to an increase of the complexity $J$ (when $N$ 
is constant) as to an increase of $N$ when the complexity $J$ is the same.
\begin{figure}[!t]
\begin{center}
\includegraphics*[width=0.49\textwidth]{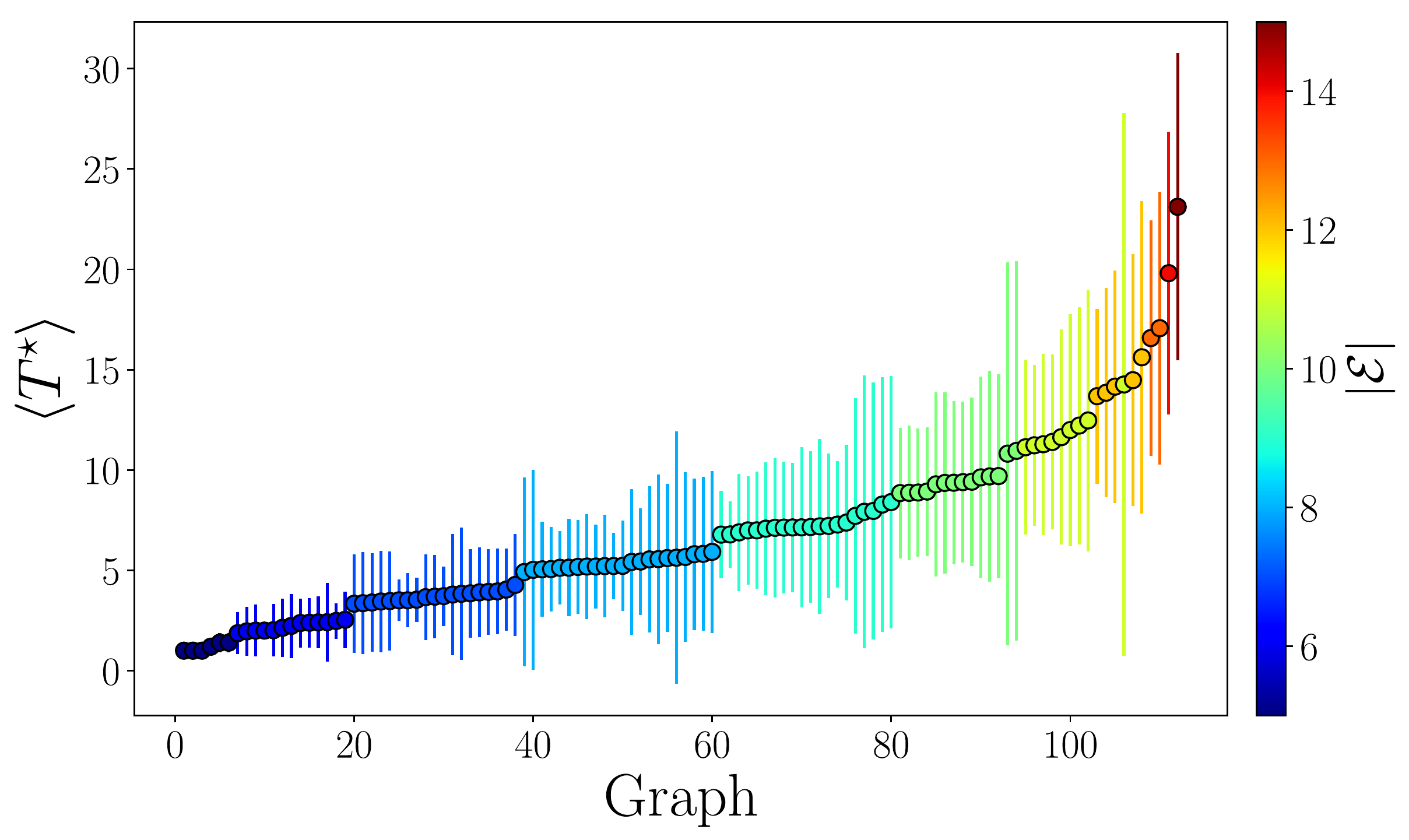} 
\end{center}
\vspace{-5mm}
\caption{\label{Fig_7} (Color online) Average lifetime $\langle T^{\star} \rangle$ of nonisomorphic connected networks with $N=6$ nodes. We analyze the times $T^\star$  for the networks presented in Fig. \ref{Fig_6} for $10^5$ Monte Carlo realizations of the cumulative damage process with $\alpha=2$ and  $\mathcal{F}^{\star}=0.75$. The error bars represent the values of the standard deviation $\sigma_{T^\star}$ of the times $T^\star$.  The different colors codified in the color bar show the number of edges $|\mathcal{E}|$ in each network.}
\end{figure}
\\[2mm]
Finally, we present some results that help us to have a graphical picture of the existing connection between the average lifetime $\langle T^{\star}\rangle$ and the topology of a network. In Figs. \ref{Fig_6} and \ref{Fig_7}, we analyze all the different undirected connected graphs with $N=6$ nodes. We study 112 different connected graphs without considering isomorphic structures; this catalog of graphs was obtained from Ref. \cite{ConnectedGraphs}. We observe that, in some cases, the system is extremely fragile and just one hit can reduce abruptly the functionality; for this reason, we analyze $T^{\star}$ with a high value of the threshold functionality $\mathcal{F}^\star$. By using the values $\alpha=2$ and $\mathcal{F}^\star=0.75$, we calculate the ensemble average of the lifespans $T^{\star}$ obtained from $10^5$ realizations of our cumulative damage algorithm. In Fig. \ref{Fig_7} we present the numerical values of $\langle T^{\star}\rangle$ and the respective standard deviation  $\sigma_{T^\star}$ as an error bar, and we sort the 
graphs in terms of the value $\left \langle T^{\star} \right \rangle$. In Fig. \ref{Fig_6} we show all the structures analyzed from the less stable under damage to the more resilient structures that live longer. In Figs. \ref{Fig_6} and \ref{Fig_7}, we see that the structures with the lowest values of $\left \langle T^{\star} \right \rangle$  are the trees with $|\mathcal{E}|=5$ depicted in the graphs 1 to 6. In trees, the failure in a single connection compromises the complete graph and for  $\alpha=2$ and $\mathcal{F}^\star=0.75$ we observe that one or two hits are sufficient to kill the system. In graphs 7 to 19, we have $|\mathcal{E}|=6$ and now all these structures have one cycle (a closed path with three or more nodes on the network that starts and end in the same node \cite{west_introduction_2000}). With the addition of a new line, we have the configurations with $|\mathcal{E}|=7$ in graphs 20-38, all these graphs have two cycles (for example, two triangles in graphs 20-24,28,29, and 38). 
The gradual increase of the number of lines allows having more cycles of different sizes; 
for example, two triangles and one square in configuration 42. 
In this way, the networks with more connections can have several cycles making them 
resilient to the damage of a particular line 
since there are many different alternatives to maintain operational conditions in the transport. 
The fully connected graph 112 is the configuration that lives longer; however, in several 
applications and real-world systems, 
each link in a system has a cost (for instance, consider the increase of development time in 
living beings with increasing `complexity'). As a consequence, it is 
important to know the `best way' to organize a determined number 
of links to maximize the survival probabilities to achieve the maximal longevity 
$\left\langle T^\star \right\rangle$. 
In addition to these results, the standard deviation $\sigma_{T^\star}$ reflects different ways that cumulative damage of the system can be distributed in the lines. 
\section{Conclusions}
In this paper, we explore the concept of aging as a consequence of cumulative 
random damage and imperfect reparation (misrepair) in a complex system. We model this phenomenon as a dynamical 
process on weighted networks. The formalism introduced includes three characteristics: (1) 
an algorithm to produce preferential random damage on the connections of the 
network that evolves with time concentrating the damage in particular parts,
(2) a collective task that requires communication between all the elements of the network and, 
(3) a global measure that quantifies the performance of the structure in a determined configuration. 
With these features, we analyze the evolution of the transport capacity in networks and how 
the global capacity of the system is gradually reduced as a consequence of the accumulation 
of faults (misrepairs). 
This process of gradual deterioration of the functionality of the structure is conceived as `aging' of the complex system.
Moreover, we define a threshold value of the functionality that determines when the system is alive. 
Through Monte Carlo simulations of this algorithm, we analyze statistically the process of 
aging and the lifespan in different types of networks. 
We explore how the structure of the system influences its longevity. 
The results reveal that complex structures are more resilient to cumulative 
damage and as a consequence live longer. 
The methods introduced with this model are general and can be used to analyze the 
global effects of cumulative random damage in different processes modeled by networks. A subject 
of further interest could be the analysis of the relationship of the complexity of 
systems and the time dependence and evolutions of
their survival probabilities. 
\section{Appendices}
\subsection{Asymptotic fault distribution}
\label{Section_appendixA}
The goal of this part is to analyze the time evolution of the fault number distribution 
defined by Eq. (\ref{problinks}) and how this behavior is related to the size of the structure.
To this end let us first consider the probability per time increment $\Delta T$ 
that a fault is generated anywhere in the whole structure. 
This probability is with Eq. (\ref{problinks}) given by
\begin{equation}
 \label{totalprob}
 \sum_{(l,m) \in \mathcal{E}} \pi_{lm}(T) =\frac{ \sum_{(i,j) 
 \in \mathcal{E}} h_{ij}(T)}{\sum_{(l,m) \in \mathcal{E}} h_{lm}(T)} \\
= 1,
\end{equation}
a relation that reflects our initial assumption that 
in each time increment $\Delta T$ a new fault
is generated, thus the probability of occurrence of one fault somewhere 
in the set $\mathcal{E}$ in the time increment $\Delta T$ is one. 
Therefore, the total number of faults generated in the structure (in the set $\mathcal{E}$) within the time interval
$[0,T]$ is given by
\begin{equation}
 \label{dgeneration}
 \sum_{(l,m) \in \mathcal{E}} (h_{lm}(T) - h_{lm}(0)) =  T,
\end{equation}
where $\sum_{(l,m) \in \mathcal{E}} h_{lm}(0) = |\mathcal{E}|$ with $h_{lm}(0)=1$.
On the other hand, we have in Eq. (\ref{problinks}) the sum
\begin{equation}
 \label{normalization}
 \sum_{(l,m) \in \mathcal{E}} h_{lm}(T) = \sum_{(l,m) \in \mathcal{E}} h_{lm}(0)  + T = |\mathcal{E}| +T,
\end{equation}
where the number of lines in the structure is
$|\mathcal{E}| =\frac{1}{2} \sum_{i=1}^N\sum_{j=1}^N A_{ij} = \frac{1}{2}\sum_{i=1}^N k_i$. 
The number of faults $h_{ij}(T)-1$ in each line is bounded $0 \leq h_{ij}(T)-1 \leq T$ by the total number $T$ of faults in the structure.
Then Eq. (\ref{problinks}) can be written as
\begin{equation}
 \label{write1as}
 \pi_{ij}(T) = \frac{h_{ij}(T-1)}{\sum_{(l,m) \in \mathcal{E}} h_{lm}(T-1) } = \frac{h_{ij}(T-1)}{|\mathcal{E}|+T-1} .
\end{equation}
Now we introduce the number of lines $N(h,T)$ that have $h-1$ faults at time $T$.
Then the total number $|\mathcal{E}|$ of lines can be represented as
\begin{equation}
\label{wehave}
\sum_{h=1}^{T+1} N(h,T) = |\mathcal{E}|,
\end{equation}
and we can write Eq. (\ref{normalization}) in the form
\begin{equation}
\label{andfurther}
\sum_{h=1}^{T+1} N(h,T) h = |\mathcal{E}| + T .
\end{equation}
It is instructive to consider the fraction $P(h,T)= \frac{N(h,T)}{|\mathcal{E}|}$ 
of lines that have $h-1$ faults at time $T$. This quantity can be identified for $|\mathcal{E}| \gg 1$
as the probability that a line $(i,j)$ has $h-1$ faults. Then relation (\ref{wehave}) becomes 
a normalization condition
\begin{equation}
\label{wehave2}
\sum_{h=1}^{T+1} P(h,T) \approx \frac{1}{\Delta h}\int_{1}^{T+1}P(h,T){\rm d}h  = 1 ,
\end{equation}
with $P(h,T=0)= \delta_{h1}$. Therefore, from Eq. (\ref{andfurther}), we get $\sum_{h=1}^{T+1} P(h,T) h = \frac{T+|\mathcal{E}|}{|\mathcal{E}|}$, where the initial condition reflects that at $T=0$ all lines have no faults, and we have $0\leq h-1 \leq T$.
\\[2mm]
Now let us consider the time evolution of $P(h,T)$. From Eq. (\ref{problinks}) follows the master equation 
\begin{multline}
\label{mastereq}
P(h,T)-P(h,T-1)=\frac{1}{T+|\mathcal{E}|-1}\times \\ \left[ (h-1)P(h-1,T-1)
-h P(h,T-1)\right].
\end{multline}
In the following, it is convenient to introduce the continuous variables $\xi = h \delta \xi$ ($\delta \xi \to 0$) 
and $\chi = T \delta \chi$ ($\delta \chi \to 0$) 
where $\xi$ and $\chi$ are kept constant when $\delta \chi,\delta \xi \to 0$. 
The continuous variables $\xi$ and $\chi$ can be conceived as continuous fault- and time measures, respectively.
In this sense the analysis to follow covers the asymptotic behavior for
$h=\frac{\xi}{\delta \xi} \gg 1$ and $T=\frac{\chi}{\delta \chi} \gg 1$. 
We have to be aware that this asymptotic behavior is a pure intrinsic 
property of the dynamics of Eq. (\ref{problinks}) 
where $T$ may exceed the longevity of the complex system represented by a network.
Let us introduce the probability density $p(\xi,\chi)$ that for a line the fault variable has value $\xi$, namely,
\begin{equation}
 \label{newfields}
 p(\xi,\chi) \delta \xi  = P(h,T);
\end{equation}
thus $P(h,T)-P(h,T-1) = \delta \xi (p(\xi,\chi)-p(\xi,\chi-\delta \chi)) \approx \delta \xi \delta \chi \frac{\partial}{\partial \chi}p(\xi,\chi)$. The normalization then is expressed as
$\sum_{h=1}^{T+1}P(h,T) \to \int_{0}^{\chi} p(\xi,\chi){\rm d}\xi =1$. From relation (\ref{newfields}) follows that
this approach becomes pertinent especially for `large' line numbers $|\mathcal{E}|$.
Then the master equation (\ref{mastereq}) can be written for the density (\ref{newfields}) as
\begin{multline}
 \label{newastereq}
 \frac{\partial}{\partial \chi}p(\xi,\chi)= - \frac{1}{ \delta \chi (T+|\mathcal{E}|-1)} \frac{1}{\delta \xi}\times\\
\left((\xi-\delta \xi)p(\xi-\delta\xi,\chi-\delta \chi)-\xi p(\xi,\chi-\delta \chi)\right) ,
\end{multline}
where $\delta \chi, \delta \xi \to 0$. Now, by using $\delta \chi(T+|\mathcal{E}|-1) = \chi +(\mathcal{E}|-1)\delta \chi \to \chi$ and $(h-1)\delta \xi =\xi-\delta \xi \to \xi$ and with \\ \\
$\lim_{\delta \xi\rightarrow 0} \frac{1}{\delta \xi}\left((\xi-\delta \xi) p(\xi-\delta\xi,\chi-\delta \chi)-\xi p(\xi,\chi-\delta \chi)\right) = 
-\frac{\partial }{\partial \xi}(\xi p(\xi,\chi))$, Eq. (\ref{newastereq}) takes the representation
\begin{equation}
 \label{masterequationcontnuum}
 \frac{\partial }{\partial \chi} p(\xi,\chi) = -\frac{1}{\chi}\frac{\partial }{\partial \xi}\left(\xi\, p(\xi,\chi)\right),
\end{equation}
where $p(\xi,\chi)$ fulfills the normalization [reflecting Eq. (\ref{wehave})]
\begin{equation}
 \label{normalizationp-distr}
 \int_0^{\chi} p(\xi,\chi){\rm d}\xi = 1,
\end{equation}
and the relation corresponding to Eqs. (\ref{normalization}) and (\ref{andfurther}) writes 
\begin{equation}
 \label{thateqwritesasfollows}
 \int_0^{\chi} p(\xi,\chi) \xi {\rm d}\xi = \lim_{\delta \xi \to 0} \frac{T \delta \chi + |\mathcal{E}|\delta \chi) }{|\mathcal{E}|} = \frac{\chi}{|\mathcal{E}|},
\end{equation}
where we assume $\frac{\delta \xi}{\delta \chi}=1$. 
The general solution of Eq. (\ref{masterequationcontnuum}) can be obtained by the separation
ansatz
\begin{equation}
 \label{separation}
 p(\xi,\chi) = u(\xi)v(T) ,
\end{equation}
which leads to 
\begin{equation}
 \label{sepaeq}
 \frac{\chi}{v(\chi)}  \frac{\partial }{\partial \chi} v(\chi) = - \frac{1}{u(\xi)}\frac{\partial }{\partial \xi}\left(\xi\, u(\chi)\right) = -\lambda .
\end{equation}
Thus we get 
\begin{equation}
\label{eig1eq}
\frac{\partial }{\partial \chi} v(\chi) = -\frac{\lambda}{\chi} v(\chi)
\end{equation}
and
\begin{equation}
\label{eig2eq}
 \frac{\partial }{\partial \xi}\left(\xi\, u(\chi)\right) = \lambda u(\xi) .
\end{equation}
These two equations are solved by $u(\xi)= C_1 \xi^{\lambda-1}$ and $v(T)= C_2\chi^{-\lambda}$; thus we obtain for Eq. (\ref{separation})
\begin{equation}
 \label{wehavthen}
 p(\xi,\chi) = C \frac{\xi^{\lambda-1}}{\chi^{\lambda}},
\end{equation}
where the two constants $C$ and $\lambda$ are to be determined. From the normalization condition in Eq. 
(\ref{normalizationp-distr}) follows that $C=\lambda$ thus the normalized density has the
representation
\begin{equation}
 \label{normaden}
 p(\xi,\chi) =\frac{\lambda \xi^{\lambda-1}}{\chi^{\lambda}} , \hspace{1cm} 0\leq \xi \leq \chi , \hspace{1cm} \chi \geq 0 .
\end{equation}
Finally, $\lambda$ is determined from condition (\ref{thateqwritesasfollows}) to arrive at
\begin{equation}
\label{finaldet}
 \lambda \chi^{-\lambda}  \int_0^{\chi} \xi^{\lambda-1}\xi {\rm d}\xi = \frac{\lambda \chi^{-\lambda}\chi^{\lambda+1}}{\lambda+1} = \frac{\lambda}{\lambda+1}\chi = \frac{\chi}{|\mathcal{E}|} ,
\end{equation}
which yields
\begin{equation}
 \label{eignaval}
 \lambda(|\mathcal{E}|) = \frac{1}{|\mathcal{E}|-1} \approx \frac{1}{|\mathcal{E}|} ,\hspace{1cm} |\mathcal{E}| \gg 1 .
\end{equation}
We observe that $0< \lambda(|\mathcal{E}|) < 1$ thus $\xi^{\lambda -1}$ ($-1<\lambda -1 < 0$).
So the density (\ref{normaden}) is finally obtained as
\begin{equation}
\label{stationaryp}
p(\xi,\chi) = \frac{1}{|\mathcal{E}|-1}\frac{\xi^{ \frac{1}{|\mathcal{E}|-1} -1}}{\chi^{\frac{1}{|\mathcal{E}|-1}}} \, \Theta(\chi-\xi),
\end{equation}
where we introduced the Heaviside step function, defined by $\Theta(u)=1$ for $u\geq 0$ and $\Theta(u)=0$ for $u<0$, to indicate that $p(\xi,\chi) \neq 0$ is only nonvanishing for 
$0\leq \xi \leq \chi$ ($\chi \geq 0$).
Equation (\ref{stationaryp}) represents the asymptotic distribution that develops 
by the dynamics of fault evolution in Eq. (\ref{problinks}) and is the main result of this paragraph.
As mentioned above the present analysis is especially pertinent for large networks with $|\mathcal{E}|\gg 1$. 
In these cases
we may put $\lambda(|\mathcal{E}|) \approx |\mathcal{E}|^{-1}$ in the density of Eq. (\ref{stationaryp}).
\\[2mm]
The inverse power-law scaling of the density (\ref{stationaryp}) as limiting distribution of the preferential 
fault generation
strategy of Eq. (\ref{problinks}) for a fixed finite time $\chi$ exhibits 
a large number of links with small fault numbers and
a small number of links with very large fault numbers. The smaller $|\mathcal{E}|$, 
the slower the density (\ref{stationaryp})
is decaying so that the incidence of high fault numbers becomes larger. On the other hand, in very large 
structures $|\mathcal{E}| \to \infty$, the density (\ref{stationaryp}) becomes concentrated 
around $\xi=0$ where low fault numbers have extremely high incidence 
whereas high fault numbers extremely low incidence. 
To show these size effects more closely a further instructive quantity is 
the cumulative probability that the fault measure does not exceed a certain value
$\xi_0 \leq \chi$ at time
$\chi$. This cumulative probability which we denote as ${\cal P}(\xi \leq \xi_0,\chi)$ yields
\begin{equation}
 \label{cumulatveprob}
 {\cal P}(\xi \leq \xi_0,\chi) = \int_0^{\xi_0} p(\xi,\chi){\rm d}\xi = 
 \left(\frac{\xi_0}{\chi}\right)^{\frac{1}{|\mathcal{E}|-1}} ,\, \xi_0 \leq \chi,
\end{equation}
where for $\xi_0\rightarrow 0$ this quantity is vanishing which reflects the initial condition that at $\chi=0$ 
there are no faults in the structure and $0 \leq {\cal P}(\xi \leq \xi_0,\chi) \leq 1$.
When the number of lines $|\mathcal{E}|$ increases, the exponent $\frac{1}{|\mathcal{E}|}\rightarrow 0$, and thus distribution 
(\ref{cumulatveprob}) for $\xi >0$ `immediately' approaches one. This universal behavior becomes extremely 
pronounced in
the limit $|\mathcal{E}|\rightarrow \infty$ of infinite networks where we obtain for Eq. (\ref{stationaryp}) 
the distributional relation
\begin{align}
\nonumber
 &\lim_{|\mathcal{E}| \to \infty} p(\xi,\chi) = \lim_{|\mathcal{E}| \to \infty}\frac{1}{|\mathcal{E}|} \frac{\xi^{\frac{1}{|\mathcal{E}|}-1}}{\chi^{\frac{1}{|\mathcal{E}|}}}\\ 
&= 
\lim_{|\mathcal{E}| \to \infty}\chi^{-\frac{1}{|\mathcal{E}|}} \frac{d}{d\xi} \xi^{\frac{1}{|\mathcal{E}|}} =\frac{d}{d\xi}\Theta(\xi)=\delta(\xi),
 \label{limit}
\end{align}
where $\delta(\xi)$ denotes Dirac's $\delta$-function.
The cumulative probability (\ref{cumulatveprob}) takes, in the limit of infinitely large structures,
a Heaviside step function shape, namely
$\lim_{|\mathcal{E}| \to \infty} {\cal P}(\xi \leq \xi_0,\chi) =\Theta(\xi_0)$.
In large structures $|\mathcal{E}| \rightarrow \infty $ for a fixed time `almost all edges' 
exhibit extremely small fault measures. This size effect is reflected by the fact that $p(\xi,\chi)$ 
becomes extremely concentrated around $\xi=0$ [approaching for $|\mathcal{E}|\rightarrow\infty$ a
Dirac $\delta(\xi)$-function
shape], thus (\ref{cumulatveprob}) at small $\xi$ immediately jumps to one. 
\subsection{Random walk characterization}
\label{Section_appendixB}
In this appendix, we derive briefly some basic random walk 
quantities utilized. There are two different timescales relevant in our model. 
The random walk which we assume to 
`simulate' the life-maintaining functions is much faster than the dynamics of the aging process. 
Let $t \approx \tau(T)$ be the global time of Eq. (\ref{tauglobal}) that defines a characteristic time
scale of the random walk. Then we have $\tau(T) \ll \Delta T$, i.e. the characteristic 
timescale $\Delta T$
where aging changes occur is much larger than the timescale of motions 
of the random walker on the network. In other words, the dynamics of aging changes are much slower than
the motion of the random walker.
We assume that the complex system is ``alive'' if any two nodes of the network
can exchange information in a sufficiently short time, i.e. if the global time $\tau(T)$  
does not exceed a certain critical value [see Eqs. (\ref{tauglobal}) and (\ref{TauiSpect}) and the functionality defined in Eq. (\ref{F_ratioT})]. We assume a {\it Markovian} 
time discrete random walker that performs at any time increment $\Delta t$ a random step from one node to another.
This process is defined by the master equation \cite{Hughes,NohRieger2004,LambiottePRE2011}
\begin{equation}
 \label{mastereqnNRW}
 P_{ij}(t+\Delta t,T) = \sum_{\ell=1}^NP_{i\ell}(t,T)w_{\ell\to j}(T) 
\end{equation}
which is valid for $t \ll \Delta T =1$. In this master equation $P_{ij}(t,T)$ indicates the probability that the walker that starts its walk at node $i$ at $t=0$ occupies node $j$ at the $n$-th time step $t=n\Delta t$.
The elements $w_{i\to j}(T)$ of the one-step transition matrix 
$\mathbf{W}(T)$  represent the probabilities to hop from node $i$ to $j$ in one time increment $\Delta t$. 
Note that in general the 
transition probability matrix is not symmetric and given by [see Eqs. (\ref{OmegaijT}) and (\ref{transitionPij})] 
\begin{equation}
 \label{one-steptransmat}
 w_{i\to j}(T) = A_{ij}\frac{(h_{ij}(T))^{-\alpha}}{\Omega_i(T)},
\end{equation}
where $\Omega_i(T)=\sum_{s=1}^N A_{is}(h_{is}(T))^{-\alpha}$ denotes the weighted degree. The one-step transition matrix in Eq. (\ref{one-steptransmat}) is constructed such (due to $w_{i\to i}(T)=0$) that the walker has to change the node at any step.
The canonic representation of the ($t=n\Delta t$) of the $n$-step transition matrix is
\begin{align}\nonumber
 {\mathbf{P}}(n\Delta t,T) &= {\mathbf W}^n(T) \\
&=
\sum_{m=1}^N (\lambda_m(T))^n |\phi_m(T)\rangle\langle {\bar \phi}_m(T)| .
 \label{mastereqnNRWtimeevo}
\end{align}
We use Dirac's (bra-ket) notation. In Eq. (\ref{mastereqnNRWtimeevo}), $|\phi_m(T)\rangle, \langle {\bar \phi}_m(T)|$ denote, respectively, the right- and left  eigenvectors of the transition matrix with the respective eigenvalues $-1 \leq \lambda_m(T) \leq 1$. The walk which we assume to take place on an undirected connected and finite network ($N<\infty$) corresponds to an {\it aperiodic ergodic} Markov chain with 
the unique eigenvalue $\lambda_1(T) =1 \forall T $ reflecting row stochasticity (with corresponding right-eigenvector having identical components)
of the transition matrix $\sum_{j=1}^N w_{i \to j}(T) = 1 $ and $|\lambda_m(T)| \leq 1$ maintained $ \forall T$ for $m=1,\ldots ,N$ (see Ref. \cite{FractionalBook2019} for a detailed analysis).
The stationary distribution $ P_{j}^{(\infty)}(T)\equiv \lim_{t\to\infty}P_{ij}(t,T) $, which gives the probability to find the random walker in the node $j$ in the limit $t\to\infty$, is given by \cite{RiascosMateos2012,MasudaPhysRep2017,FractionalBook2019}
\begin{equation}\label{stationary}
P_{j}^{(\infty)}(T) = \langle i|\phi_1(T)\rangle\langle \bar \phi_1(T)|j\rangle= 
\frac{\Omega_{j}(T)}{\sum_{l=1}^N \Omega_l(T)} 
\end{equation}
where, since $\left\langle i|\phi_1(T)\right\rangle=\mathrm{constant}$, the stationary distribution $P_j^{(\infty)}(T)$ does not depend on the initial condition. 
\\[2mm]
Additionally, we have the mean first passage time $\left\langle {\cal T}_{ij}\right\rangle $ that gives the average number of time steps (in units of $\Delta t$) the walker needs 
to travel from node $i$ to node $j$ in the form (see Refs. \cite{NohRieger2004,RiascosMateos2012,ZhangPRE2013,FractionalBook2019} for a complete derivation)
\begin{multline}
\label{TijSpect}
\left\langle {\cal T}_{ij}\right\rangle = 
 \frac{\delta_{ij}}{{\left\langle j|\phi_1(T)\right\rangle \left\langle\bar{\phi}_1(T)|j\right\rangle}} \\
 +\sum_{\ell=2}^N \frac{\left\langle j|\phi_\ell(T)\right\rangle \left\langle\bar{\phi}_\ell(T)|j\right\rangle-\left
 \langle i|\phi_\ell(T)\right\rangle \left\langle\bar{\phi}_\ell(T)|j\right\rangle}{(1-\lambda_\ell(T))\left\langle j|\phi_1(T)\right\rangle \left\langle\bar{\phi}_1(T)|j\right\rangle}.
\end{multline}
In this relation for $i=j$ the second term and for $j\neq i$ the first term vanishes.
For $i=j$ this relation gives
the {\it mean first return time} or {\it mean recurrence time} \cite{Kac1947,FractionalBook2019}
\begin{equation}
\label{meanrcurrencetime}
\left\langle {\cal T}_{jj}\right\rangle =\frac{1}{{\left\langle j|\phi_1(T)\right\rangle \left\langle\bar{\phi}_1(T)|j\right\rangle}}.
\end{equation}
This is the {\it Kac-formula} relating the mean recurrence time with the inverse of the
stationary distribution $P_{j}^{(\infty)}(T)$. 
\\[2mm]
On the other hand,  we can define the characteristic time 
\begin{equation}\label{timetau_2}
    \tau_j(T)=\sum_{l=2}^N\frac{1}{1-\lambda_l(T)}\frac{\left\langle j|\phi_l(T)\right\rangle \left\langle\bar{\phi}_l(T)|j\right\rangle}{\left\langle j|\phi_1(T)\right\rangle \left\langle\bar{\phi}_1(T)|j\right\rangle}\, ,
\end{equation}
and by considering the relation in Eq. (\ref{TijSpect}), we have \cite{NohRieger2004}
\begin{equation}
	\langle  {\cal T}_{ij} \rangle-\langle  {\cal T}_{ji}\rangle=\tau_j-\tau_i.
\end{equation}
This result describes the asymmetry of transport on networks. The quantity $C_j=1/\tau_j$ is the random 
walk centrality introduced in Ref.  \cite{NohRieger2004}. In contrast with other centrality measures, 
defined to describe characteristics associated with the topology of the network, $ C_j $ 
is a quantity that describes the capacity of a random walker to reach the node $ j $. 
The random walker reaches nodes $j$ with higher centrality $C_j$ more easily. On the other hand, $\tau_j $ gives a value related with the average number of steps needed to reach node $ j $ from any node in the network \cite{NohRieger2004}. 
\\[2mm]
Now we can define two kinds of global times. The first one is an estimate of the average time to reach any node {\it different from the departure node}
which yields with Eq. (\ref{timetau_2})
\begin{equation}
 \label{eq4globaltime}
 \tau(T)= \frac{1}{N}\sum_{j=1}^N \tau_j(T).
\end{equation}
Furthermore, the mean recurrence return time in Eq. (\ref{meanrcurrencetime}) {\it averaged} over all nodes defines a further global measure for the speed of the random walk.
This quantity is obtained as
\begin{equation}
 \label{averagretrutime}
 {\cal T}(T) = \frac{1}{N} \sum_{i=1}^N  \left\langle {\cal T}_{ii}\right\rangle=\frac{1}{N} \sum_{i=1}^N \frac{1}{P_{i}^{(\infty)}(T)} .
 \end{equation}
In our study, we define the functionality of the system in Eq. (\ref{F_ratioT}) by 
using the global time $\tau(T)$ of Eq. (\ref{eq4globaltime}) to quantify the capacity of the random walker to explore a network in a given configuration of the system at time $T$. The value  $\tau(T)$ includes important information of the process since it considers the eigenvalues and eigenvectors of the respective transition matrix. Several studies have been shown that this global time is a good measure of the capacity of a random walker to explore a network \cite{RiascosMateos2012,RiascosMichelitsch2017_gL,FractionalBook2019}.  Finally, the time ${\cal T}(T)$ has the computational advantage that the eigenvalues and eigenvectors of the one-step transition matrix in Eq. (\ref{one-steptransmat}) do not need to be determined and can be used to define a simplified version of the functionality. 
%
%

\begin{thebibliography}{33}%
\makeatletter
\providecommand \@ifxundefined [1]{%
 \@ifx{#1\undefined}
}%
\providecommand \@ifnum [1]{%
 \ifnum #1\expandafter \@firstoftwo
 \else \expandafter \@secondoftwo
 \fi
}%
\providecommand \@ifx [1]{%
 \ifx #1\expandafter \@firstoftwo
 \else \expandafter \@secondoftwo
 \fi
}%
\providecommand \natexlab [1]{#1}%
\providecommand \enquote  [1]{``#1''}%
\providecommand \bibnamefont  [1]{#1}%
\providecommand \bibfnamefont [1]{#1}%
\providecommand \citenamefont [1]{#1}%
\providecommand \href@noop [0]{\@secondoftwo}%
\providecommand \href [0]{\begingroup \@sanitize@url \@href}%
\providecommand \@href[1]{\@@startlink{#1}\@@href}%
\providecommand \@@href[1]{\endgroup#1\@@endlink}%
\providecommand \@sanitize@url [0]{\catcode `\\12\catcode `\$12\catcode
  `\&12\catcode `\#12\catcode `\^12\catcode `\_12\catcode `\%12\relax}%
\providecommand \@@startlink[1]{}%
\providecommand \@@endlink[0]{}%
\providecommand \url  [0]{\begingroup\@sanitize@url \@url }%
\providecommand \@url [1]{\endgroup\@href {#1}{\urlprefix }}%
\providecommand \urlprefix  [0]{URL }%
\providecommand \Eprint [0]{\href }%
\providecommand \doibase [0]{http://dx.doi.org/}%
\providecommand \selectlanguage [0]{\@gobble}%
\providecommand \bibinfo  [0]{\@secondoftwo}%
\providecommand \bibfield  [0]{\@secondoftwo}%
\providecommand \translation [1]{[#1]}%
\providecommand \BibitemOpen [0]{}%
\providecommand \bibitemStop [0]{}%
\providecommand \bibitemNoStop [0]{.\EOS\space}%
\providecommand \EOS [0]{\spacefactor3000\relax}%
\providecommand \BibitemShut  [1]{\csname bibitem#1\endcsname}%
\let\auto@bib@innerbib\@empty
\bibitem [{\citenamefont {Kirkwood}(2005)}]{Kirkwood2005}%
  \BibitemOpen
  \bibfield  {author} {\bibinfo {author} {\bibfnamefont {T.~B.}\ \bibnamefont
  {Kirkwood}},\ }\href {\doibase https://doi.org/10.1016/j.cell.2005.01.027}
  {\bibfield  {journal} {\bibinfo  {journal} {Cell}\ }\textbf {\bibinfo
  {volume} {120}},\ \bibinfo {pages} {437 } (\bibinfo {year}
  {2005})}\BibitemShut {NoStop}%
\bibitem [{\citenamefont {L\'opez-Ot\'in}\ \emph {et~al.}(2013)\citenamefont
  {L\'opez-Ot\'in}, \citenamefont {Blasco}, \citenamefont {Partridge},
  \citenamefont {Serrano},\ and\ \citenamefont {Kroemer}}]{LopezOtin2013}%
  \BibitemOpen
  \bibfield  {author} {\bibinfo {author} {\bibfnamefont {C.}~\bibnamefont
  {L\'opez-Ot\'in}}, \bibinfo {author} {\bibfnamefont {M.~A.}\ \bibnamefont
  {Blasco}}, \bibinfo {author} {\bibfnamefont {L.}~\bibnamefont {Partridge}},
  \bibinfo {author} {\bibfnamefont {M.}~\bibnamefont {Serrano}}, \ and\
  \bibinfo {author} {\bibfnamefont {G.}~\bibnamefont {Kroemer}},\ }\href
  {\doibase https://doi.org/10.1016/j.cell.2013.05.039} {\bibfield  {journal}
  {\bibinfo  {journal} {Cell}\ }\textbf {\bibinfo {volume} {153}},\ \bibinfo
  {pages} {1194 } (\bibinfo {year} {2013})}\BibitemShut {NoStop}%
\bibitem [{\citenamefont {Cohen}(2016)}]{Cohen2016}%
  \BibitemOpen
  \bibfield  {author} {\bibinfo {author} {\bibfnamefont {A.~A.}\ \bibnamefont
  {Cohen}},\ }\href {\doibase 10.1007/s10522-015-9584-x} {\bibfield  {journal}
  {\bibinfo  {journal} {Biogerontology}\ }\textbf {\bibinfo {volume} {17}},\
  \bibinfo {pages} {205} (\bibinfo {year} {2016})}\BibitemShut {NoStop}%
\bibitem [{\citenamefont {Farrell}\ \emph {et~al.}(2016)\citenamefont
  {Farrell}, \citenamefont {Mitnitski}, \citenamefont {Rockwood},\ and\
  \citenamefont {Rutenberg}}]{Farrell2016}%
  \BibitemOpen
  \bibfield  {author} {\bibinfo {author} {\bibfnamefont {S.~G.}\ \bibnamefont
  {Farrell}}, \bibinfo {author} {\bibfnamefont {A.~B.}\ \bibnamefont
  {Mitnitski}}, \bibinfo {author} {\bibfnamefont {K.}~\bibnamefont {Rockwood}},
  \ and\ \bibinfo {author} {\bibfnamefont {A.~D.}\ \bibnamefont {Rutenberg}},\
  }\href {\doibase 10.1103/PhysRevE.94.052409} {\bibfield  {journal} {\bibinfo
  {journal} {Phys. Rev. E}\ }\textbf {\bibinfo {volume} {94}},\ \bibinfo
  {pages} {052409} (\bibinfo {year} {2016})}\BibitemShut {NoStop}%
\bibitem [{\citenamefont {Taneja}\ \emph {et~al.}(2016)\citenamefont {Taneja},
  \citenamefont {Mitnitski}, \citenamefont {Rockwood},\ and\ \citenamefont
  {Rutenberg}}]{Taneja2016}%
  \BibitemOpen
  \bibfield  {author} {\bibinfo {author} {\bibfnamefont {S.}~\bibnamefont
  {Taneja}}, \bibinfo {author} {\bibfnamefont {A.~B.}\ \bibnamefont
  {Mitnitski}}, \bibinfo {author} {\bibfnamefont {K.}~\bibnamefont {Rockwood}},
  \ and\ \bibinfo {author} {\bibfnamefont {A.~D.}\ \bibnamefont {Rutenberg}},\
  }\href {\doibase 10.1103/PhysRevE.93.022309} {\bibfield  {journal} {\bibinfo
  {journal} {Phys. Rev. E}\ }\textbf {\bibinfo {volume} {93}},\ \bibinfo
  {pages} {022309} (\bibinfo {year} {2016})}\BibitemShut {NoStop}%
\bibitem [{\citenamefont {Zheng}\ \emph {et~al.}(2018)\citenamefont {Zheng},
  \citenamefont {Cao}, \citenamefont {Vorobyeva}, \citenamefont {Manrique},
  \citenamefont {Song},\ and\ \citenamefont {Johnson}}]{Manrique2018}%
  \BibitemOpen
  \bibfield  {author} {\bibinfo {author} {\bibfnamefont {M.}~\bibnamefont
  {Zheng}}, \bibinfo {author} {\bibfnamefont {Z.}~\bibnamefont {Cao}}, \bibinfo
  {author} {\bibfnamefont {Y.}~\bibnamefont {Vorobyeva}}, \bibinfo {author}
  {\bibfnamefont {P.}~\bibnamefont {Manrique}}, \bibinfo {author}
  {\bibfnamefont {C.}~\bibnamefont {Song}}, \ and\ \bibinfo {author}
  {\bibfnamefont {N.~F.}\ \bibnamefont {Johnson}},\ }\href
  {https://doi.org/10.1038/s41598-018-22027-z} {\bibfield  {journal} {\bibinfo
  {journal} {Sci. Rep.}\ }\textbf {\bibinfo {volume} {8}},\ \bibinfo {pages}
  {3552} (\bibinfo {year} {2018})}\BibitemShut {NoStop}%
\bibitem [{\citenamefont {Hershey}(1986)}]{Hershey1986}%
  \BibitemOpen
  \bibfield  {author} {\bibinfo {author} {\bibfnamefont {D.}~\bibnamefont
  {Hershey}},\ }\href {\doibase 10.1002/sres.3850030102} {\bibfield  {journal}
  {\bibinfo  {journal} {Systems Research}\ }\textbf {\bibinfo {volume} {3}},\
  \bibinfo {pages} {3} (\bibinfo {year} {1986})}\BibitemShut {NoStop}%
\bibitem [{\citenamefont {West}(2017)}]{west2018scale}%
  \BibitemOpen
  \bibfield  {author} {\bibinfo {author} {\bibfnamefont {G.}~\bibnamefont
  {West}},\ }\href@noop {} {\emph {\bibinfo {title} {Scale: The Universal Laws
  of Growth, Innovation, Sustainability, and the Pace of Life in Organisms,
  Cities, Economies, and Companies}}}\ (\bibinfo  {publisher} {Penguin Press},\
  \bibinfo {address} {New York},\ \bibinfo {year} {2017})\BibitemShut {NoStop}%
\bibitem [{\citenamefont {Wang}\ \emph {et~al.}(2009)\citenamefont {Wang},
  \citenamefont {Michelitsch}, \citenamefont {Wunderlin},\ and\ \citenamefont
  {Mahadeva}}]{WangMiWun2009}%
  \BibitemOpen
  \bibfield  {author} {\bibinfo {author} {\bibfnamefont {J.}~\bibnamefont
  {Wang}}, \bibinfo {author} {\bibfnamefont {T.}~\bibnamefont {Michelitsch}},
  \bibinfo {author} {\bibfnamefont {A.}~\bibnamefont {Wunderlin}}, \ and\
  \bibinfo {author} {\bibfnamefont {R.}~\bibnamefont {Mahadeva}},\ }\href@noop
  {} {\enquote {\bibinfo {title} {Aging as a consequence of misrepair a novel
  theory of aging},}\ } (\bibinfo {year} {2009}),\ \Eprint
  {http://arxiv.org/abs/arXiv:0904.0575} {arXiv:0904.0575} \BibitemShut
  {NoStop}%
\bibitem [{\citenamefont {Wang-Michelitsch}\ and\ \citenamefont
  {Michelitsch}(2018)}]{WangMi2018}%
  \BibitemOpen
  \bibfield  {author} {\bibinfo {author} {\bibfnamefont {J.}~\bibnamefont
  {Wang-Michelitsch}}\ and\ \bibinfo {author} {\bibfnamefont {T.}~\bibnamefont
  {Michelitsch}},\ }\href@noop {} {\enquote {\bibinfo {title} {Potential of
  longevity: Hidden in structural complexity},}\ } (\bibinfo {year} {2018}),\
  \Eprint {http://arxiv.org/abs/arXiv:1505.03902} {arXiv:1505.03902}
  \BibitemShut {NoStop}%
\bibitem [{\citenamefont {Wang-Michelitsch}\ and\ \citenamefont
  {Michelitsch}(2015{\natexlab{a}})}]{WangMi2015}%
  \BibitemOpen
  \bibfield  {author} {\bibinfo {author} {\bibfnamefont {J.}~\bibnamefont
  {Wang-Michelitsch}}\ and\ \bibinfo {author} {\bibfnamefont {T.}~\bibnamefont
  {Michelitsch}},\ }\href@noop {} {\enquote {\bibinfo {title} {Aging as a
  process of accumulation of misrepairs},}\ } (\bibinfo {year}
  {2015}{\natexlab{a}}),\ \Eprint {http://arxiv.org/abs/arXiv:1503.07163}
  {arXiv:1503.07163} \BibitemShut {NoStop}%
\bibitem [{\citenamefont {Hughes}(1996)}]{Hughes}%
  \BibitemOpen
  \bibfield  {author} {\bibinfo {author} {\bibfnamefont {B.~D.}\ \bibnamefont
  {Hughes}},\ }\href@noop {} {\emph {\bibinfo {title} {Random Walks and Random
  Environments: Vol. 1: Random Walks}}}\ (\bibinfo  {publisher} {Oxford
  University Press},\ \bibinfo {address} {Oxford},\ \bibinfo {year}
  {1996})\BibitemShut {NoStop}%
\bibitem [{\citenamefont {Masuda}\ \emph {et~al.}(2017)\citenamefont {Masuda},
  \citenamefont {Porter},\ and\ \citenamefont {Lambiotte}}]{MasudaPhysRep2017}%
  \BibitemOpen
  \bibfield  {author} {\bibinfo {author} {\bibfnamefont {N.}~\bibnamefont
  {Masuda}}, \bibinfo {author} {\bibfnamefont {M.~A.}\ \bibnamefont {Porter}},
  \ and\ \bibinfo {author} {\bibfnamefont {R.}~\bibnamefont {Lambiotte}},\
  }\href {https://doi.org/10.1016/j.physrep.2017.07.007} {\bibfield  {journal}
  {\bibinfo  {journal} {Phys. Rep.}\ }\textbf {\bibinfo {volume} {716--717}},\
  \bibinfo {pages} {1} (\bibinfo {year} {2017})}\BibitemShut {NoStop}%
\bibitem [{\citenamefont {Arenas}\ \emph {et~al.}(2008)\citenamefont {Arenas},
  \citenamefont {D\'iaz-Guilera}, \citenamefont {Kurths}, \citenamefont
  {Moreno},\ and\ \citenamefont {Zhou}}]{Arenas2008}%
  \BibitemOpen
  \bibfield  {author} {\bibinfo {author} {\bibfnamefont {A.}~\bibnamefont
  {Arenas}}, \bibinfo {author} {\bibfnamefont {A.}~\bibnamefont
  {D\'iaz-Guilera}}, \bibinfo {author} {\bibfnamefont {J.}~\bibnamefont
  {Kurths}}, \bibinfo {author} {\bibfnamefont {Y.}~\bibnamefont {Moreno}}, \
  and\ \bibinfo {author} {\bibfnamefont {C.}~\bibnamefont {Zhou}},\ }\href
  {\doibase https://doi.org/10.1016/j.physrep.2008.09.002} {\bibfield
  {journal} {\bibinfo  {journal} {Phys. Rep.}\ }\textbf {\bibinfo {volume}
  {469}},\ \bibinfo {pages} {93 } (\bibinfo {year} {2008})}\BibitemShut
  {NoStop}%
\bibitem [{\citenamefont {Blanchard}\ and\ \citenamefont
  {Volchenkov}(2011)}]{BlanchardBook2011}%
  \BibitemOpen
  \bibfield  {author} {\bibinfo {author} {\bibfnamefont {P.}~\bibnamefont
  {Blanchard}}\ and\ \bibinfo {author} {\bibfnamefont {D.}~\bibnamefont
  {Volchenkov}},\ }\href@noop {} {\emph {\bibinfo {title} {{Random Walks and
  Diffusions on Graphs and Databases: An Introduction}}}},\ Springer {S}eries
  in Synergetics\ (\bibinfo  {publisher} {Springer},\ \bibinfo {address}
  {Berlin},\ \bibinfo {year} {2011})\BibitemShut {NoStop}%
\bibitem [{\citenamefont {Michelitsch}\ \emph {et~al.}(2019)\citenamefont
  {Michelitsch}, \citenamefont {Riascos}, \citenamefont {Collet}, \citenamefont
  {Nowakowski},\ and\ \citenamefont {Nicolleau}}]{FractionalBook2019}%
  \BibitemOpen
  \bibfield  {author} {\bibinfo {author} {\bibfnamefont {T.~M.}\ \bibnamefont
  {Michelitsch}}, \bibinfo {author} {\bibfnamefont {A.~P.}\ \bibnamefont
  {Riascos}}, \bibinfo {author} {\bibfnamefont {B.~A.}\ \bibnamefont {Collet}},
  \bibinfo {author} {\bibfnamefont {A.~F.}\ \bibnamefont {Nowakowski}}, \ and\
  \bibinfo {author} {\bibfnamefont {F.~C. G.~A.}\ \bibnamefont {Nicolleau}},\
  }\href@noop {} {\emph {\bibinfo {title} {Fractional Dynamics on Networks and
  Lattices}}}\ (\bibinfo  {publisher} {ISTE/Wiley},\ \bibinfo {address}
  {London},\ \bibinfo {year} {2019})\BibitemShut {NoStop}%
\bibitem [{\citenamefont {Barrat}\ \emph {et~al.}(2008)\citenamefont {Barrat},
  \citenamefont {Barth\'elemy},\ and\ \citenamefont {Vespignani}}]{VespiBook}%
  \BibitemOpen
  \bibfield  {author} {\bibinfo {author} {\bibfnamefont {A.}~\bibnamefont
  {Barrat}}, \bibinfo {author} {\bibfnamefont {M.}~\bibnamefont
  {Barth\'elemy}}, \ and\ \bibinfo {author} {\bibfnamefont {A.}~\bibnamefont
  {Vespignani}},\ }\href@noop {} {\emph {\bibinfo {title} {Dynamical Processes
  on Complex Networks}}}\ (\bibinfo  {publisher} {Cambridge University Press},\
  \bibinfo {address} {Cambridge},\ \bibinfo {year} {2008})\BibitemShut
  {NoStop}%
\bibitem [{\citenamefont {Wang-Michelitsch}\ and\ \citenamefont
  {Michelitsch}(2015{\natexlab{b}})}]{WangMi2015b}%
  \BibitemOpen
  \bibfield  {author} {\bibinfo {author} {\bibfnamefont {J.}~\bibnamefont
  {Wang-Michelitsch}}\ and\ \bibinfo {author} {\bibfnamefont {T.}~\bibnamefont
  {Michelitsch}},\ }\href@noop {} {\enquote {\bibinfo {title} {Development of
  aging changes: self-accelerating and inhomogeneous},}\ } (\bibinfo {year}
  {2015}{\natexlab{b}}),\ \Eprint {http://arxiv.org/abs/arXiv:1503.08076}
  {arXiv:1503.08076} \BibitemShut {NoStop}%
\bibitem [{\citenamefont {Barab\'asi}(2016)}]{NetworkScienceBook2016}%
  \BibitemOpen
  \bibfield  {author} {\bibinfo {author} {\bibfnamefont {A.-L.}\ \bibnamefont
  {Barab\'asi}},\ }\href@noop {} {\emph {\bibinfo {title} {Network science}}}\
  (\bibinfo  {publisher} {Cambridge University Press},\ \bibinfo {address}
  {Cambridge},\ \bibinfo {year} {2016})\BibitemShut {NoStop}%
\bibitem [{\citenamefont {Riascos}\ and\ \citenamefont
  {Mateos}(2012)}]{RiascosMateos2012}%
  \BibitemOpen
  \bibfield  {author} {\bibinfo {author} {\bibfnamefont {A.~P.}\ \bibnamefont
  {Riascos}}\ and\ \bibinfo {author} {\bibfnamefont {J.~L.}\ \bibnamefont
  {Mateos}},\ }\href {\doibase 10.1103/PhysRevE.86.056110} {\bibfield
  {journal} {\bibinfo  {journal} {Phys. Rev. E}\ }\textbf {\bibinfo {volume}
  {86}},\ \bibinfo {pages} {056110} (\bibinfo {year} {2012})}\BibitemShut
  {NoStop}%
\bibitem [{\citenamefont {Riascos}\ \emph {et~al.}(2018)\citenamefont
  {Riascos}, \citenamefont {Michelitsch}, \citenamefont {Collet}, \citenamefont
  {Nowakowski},\ and\ \citenamefont {Nicolleau}}]{RiascosMichelitsch2017_gL}%
  \BibitemOpen
  \bibfield  {author} {\bibinfo {author} {\bibfnamefont {A.~P.}\ \bibnamefont
  {Riascos}}, \bibinfo {author} {\bibfnamefont {T.~M.}\ \bibnamefont
  {Michelitsch}}, \bibinfo {author} {\bibfnamefont {B.~A.}\ \bibnamefont
  {Collet}}, \bibinfo {author} {\bibfnamefont {A.~F.}\ \bibnamefont
  {Nowakowski}}, \ and\ \bibinfo {author} {\bibfnamefont {F.~C. G.~A.}\
  \bibnamefont {Nicolleau}},\ }\href
  {http://stacks.iop.org/1742-5468/2018/i=4/a=043404} {\bibfield  {journal}
  {\bibinfo  {journal} {J. Stat. Mech.}\ }\textbf {\bibinfo {volume} {2018}},\
  \bibinfo {pages} {043404} (\bibinfo {year} {2018})}\BibitemShut {NoStop}%
\bibitem [{\citenamefont {Noh}\ and\ \citenamefont
  {Rieger}(2004)}]{NohRieger2004}%
  \BibitemOpen
  \bibfield  {author} {\bibinfo {author} {\bibfnamefont {J.~D.}\ \bibnamefont
  {Noh}}\ and\ \bibinfo {author} {\bibfnamefont {H.}~\bibnamefont {Rieger}},\
  }\href {\doibase 10.1103/PhysRevLett.92.118701} {\bibfield  {journal}
  {\bibinfo  {journal} {Phys. Rev. Lett.}\ }\textbf {\bibinfo {volume} {92}},\
  \bibinfo {pages} {118701} (\bibinfo {year} {2004})}\BibitemShut {NoStop}%
\bibitem [{Note1()}]{Note1}%
  \BibitemOpen
  \bibinfo {note} {See supplemental material for videos with Monte Carlo
  simulations of this algorithm for $\alpha =2$ and $T=0,1,\protect \ldots
  ,100$. Three different realizations are presented in the videos video1.avi,
  video2.avi, video3.avi, respectively.}\BibitemShut {Stop}%
\bibitem [{\citenamefont {Barab\'asi}\ and\ \citenamefont
  {Albert}(1999)}]{BarabasiAlbert1999}%
  \BibitemOpen
  \bibfield  {author} {\bibinfo {author} {\bibfnamefont {A.-L.}\ \bibnamefont
  {Barab\'asi}}\ and\ \bibinfo {author} {\bibfnamefont {R.}~\bibnamefont
  {Albert}},\ }\href {\doibase 10.1126/science.286.5439.509} {\bibfield
  {journal} {\bibinfo  {journal} {Science}\ }\textbf {\bibinfo {volume}
  {286}},\ \bibinfo {pages} {509} (\bibinfo {year} {1999})}\BibitemShut
  {NoStop}%
\bibitem [{\citenamefont {Watts}\ and\ \citenamefont
  {Strogatz}(1998)}]{WattsStrogatz1998}%
  \BibitemOpen
  \bibfield  {author} {\bibinfo {author} {\bibfnamefont {D.~J.}\ \bibnamefont
  {Watts}}\ and\ \bibinfo {author} {\bibfnamefont {S.~H.}\ \bibnamefont
  {Strogatz}},\ }\href {\doibase 10.1038/30918} {\bibfield  {journal} {\bibinfo
   {journal} {Nature (London)}\ }\textbf {\bibinfo {volume} {393}},\ \bibinfo
  {pages} {440} (\bibinfo {year} {1998})}\BibitemShut {NoStop}%
\bibitem [{\citenamefont {Erd\"{o}s}\ and\ \citenamefont
  {R\'{e}nyi}(1959)}]{ErdosRenyi1959}%
  \BibitemOpen
  \bibfield  {author} {\bibinfo {author} {\bibfnamefont {P.}~\bibnamefont
  {Erd\"{o}s}}\ and\ \bibinfo {author} {\bibfnamefont {A.}~\bibnamefont
  {R\'{e}nyi}},\ }\href@noop {} {\bibfield  {journal} {\bibinfo  {journal}
  {Publ. Math. (Debrecen)}\ }\textbf {\bibinfo {volume} {6}},\ \bibinfo {pages}
  {290} (\bibinfo {year} {1959})}\BibitemShut {NoStop}%
\bibitem [{Con()}]{ConnectedGraphs}%
  \BibitemOpen
  \href@noop {} {}\bibinfo {note}
  {\url{http://users.cecs.anu.edu.au/~bdm/data/graphs.html}}\BibitemShut
  {NoStop}%
\bibitem [{\citenamefont {Van~Mieghem}(2011)}]{VanMieghem}%
  \BibitemOpen
  \bibfield  {author} {\bibinfo {author} {\bibfnamefont {P.}~\bibnamefont
  {Van~Mieghem}},\ }\href@noop {} {\emph {\bibinfo {title} {Graph Spectra for
  Complex Networks}}}\ (\bibinfo  {publisher} {Cambridge University Press},\
  \bibinfo {address} {New York},\ \bibinfo {year} {2011})\BibitemShut {NoStop}%
\bibitem [{\citenamefont {Riascos}\ and\ \citenamefont
  {Mateos}(2015)}]{RiascosMateosFD2015}%
  \BibitemOpen
  \bibfield  {author} {\bibinfo {author} {\bibfnamefont {A.~P.}\ \bibnamefont
  {Riascos}}\ and\ \bibinfo {author} {\bibfnamefont {J.~L.}\ \bibnamefont
  {Mateos}},\ }\href {http://stacks.iop.org/1742-5468/2015/i=7/a=P07015}
  {\bibfield  {journal} {\bibinfo  {journal} {J. Stat. Mech.}\ }\textbf
  {\bibinfo {volume} {2015}},\ \bibinfo {pages} {P07015} (\bibinfo {year}
  {2015})}\BibitemShut {NoStop}%
\bibitem [{\citenamefont {West}(2001)}]{west_introduction_2000}%
  \BibitemOpen
  \bibfield  {author} {\bibinfo {author} {\bibfnamefont {D.~B.}\ \bibnamefont
  {West}},\ }\href@noop {} {\emph {\bibinfo {title} {Introduction to Graph
  Theory}}},\ \bibinfo {edition} {2nd}\ ed.\ (\bibinfo  {publisher} {Pearson
  Education},\ \bibinfo {address} {Singapore​},\ \bibinfo {year}
  {2001})\BibitemShut {NoStop}%
\bibitem [{\citenamefont {Lambiotte}\ \emph {et~al.}(2011)\citenamefont
  {Lambiotte}, \citenamefont {Sinatra}, \citenamefont {Delvenne}, \citenamefont
  {Evans}, \citenamefont {Barahona},\ and\ \citenamefont
  {Latora}}]{LambiottePRE2011}%
  \BibitemOpen
  \bibfield  {author} {\bibinfo {author} {\bibfnamefont {R.}~\bibnamefont
  {Lambiotte}}, \bibinfo {author} {\bibfnamefont {R.}~\bibnamefont {Sinatra}},
  \bibinfo {author} {\bibfnamefont {J.-C.}\ \bibnamefont {Delvenne}}, \bibinfo
  {author} {\bibfnamefont {T.~S.}\ \bibnamefont {Evans}}, \bibinfo {author}
  {\bibfnamefont {M.}~\bibnamefont {Barahona}}, \ and\ \bibinfo {author}
  {\bibfnamefont {V.}~\bibnamefont {Latora}},\ }\href {\doibase
  10.1103/PhysRevE.84.017102} {\bibfield  {journal} {\bibinfo  {journal} {Phys.
  Rev. E}\ }\textbf {\bibinfo {volume} {84}},\ \bibinfo {pages} {017102}
  (\bibinfo {year} {2011})}\BibitemShut {NoStop}%
\bibitem [{\citenamefont {Zhang}\ \emph {et~al.}(2013)\citenamefont {Zhang},
  \citenamefont {Shan},\ and\ \citenamefont {Chen}}]{ZhangPRE2013}%
  \BibitemOpen
  \bibfield  {author} {\bibinfo {author} {\bibfnamefont {Z.}~\bibnamefont
  {Zhang}}, \bibinfo {author} {\bibfnamefont {T.}~\bibnamefont {Shan}}, \ and\
  \bibinfo {author} {\bibfnamefont {G.}~\bibnamefont {Chen}},\ }\href {\doibase
  10.1103/PhysRevE.87.012112} {\bibfield  {journal} {\bibinfo  {journal} {Phys.
  Rev. E}\ }\textbf {\bibinfo {volume} {87}},\ \bibinfo {pages} {012112}
  (\bibinfo {year} {2013})}\BibitemShut {NoStop}%
\bibitem [{\citenamefont {Kac}(1947)}]{Kac1947}%
  \BibitemOpen
  \bibfield  {author} {\bibinfo {author} {\bibfnamefont {M.}~\bibnamefont
  {Kac}},\ }\href {https://projecteuclid.org:443/euclid.bams/1183511152}
  {\bibfield  {journal} {\bibinfo  {journal} {Bull. Amer. Math. Soc.}\ }\textbf
  {\bibinfo {volume} {53}},\ \bibinfo {pages} {1002} (\bibinfo {year}
  {1947})}\BibitemShut {NoStop}%
\end{thebibliography}

%

\end{document}